\documentclass[12pt, draftcls, peerreview, onecolumn]{IEEEtran}

\def\BibTeX{{\rm B\kern-.05em{\sc i\kern-.025em b}\kern-.08em
    T\kern-.1667em\lower.7ex\hbox{E}\kern-.125emX}}

\usepackage{bbm}
\usepackage{amsmath}
\usepackage{amsthm}
\usepackage{amsfonts}
\usepackage{amssymb}
\usepackage{amscd}
\usepackage{graphicx}
\usepackage{newlfont}
\usepackage{cite}
\usepackage{color}
\usepackage{multirow,tabularx}
\usepackage{ifthen}
\usepackage{enumitem}
\usepackage{times}
\usepackage{stackrel}
\usepackage{floatrow}
\usepackage{algorithm,algpseudocode}
\usepackage{epstopdf}
\usepackage[all]{xy}
\usepackage{url}
\usepackage[font = footnotesize,belowskip=-15pt,aboveskip=0pt]{caption}
\usepackage{booktabs}
\usepackage{multirow,tabularx}
\usepackage{hyphenat}
\usepackage{verbatim}
\usepackage{mathtools}
\usepackage{multicol,lipsum}
\usepackage[table]{xcolor} 
\usepackage{mathtools}
\usepackage{flexisym}
\usepackage{graphicx}
\usepackage{subfig}

\def\BibTeX{{\rm B\kern-.05em{\sc i\kern-.025em b}\kern-.08em
        T\kern-.1667em\lower.7ex\hbox{E}\kern-.125emX}}

\IEEEoverridecommandlockouts

\DeclarePairedDelimiter\abs{\lvert}{\rvert}%

\newcommand{\norm}[1]{\left\Vert#1\right\Vert}
\newcommand{\card}[1]{|#1|}

\newcommand{\Real}{\mathbb{R}}

\newcommand{\vecw}{\mathbf{w}}

\allowdisplaybreaks

\let\oldabs\abs
\def\abs{\@ifstar{\oldabs}{\oldabs*}}

\newcommand{\BW}{B}

\newcommand{\UesPerGrp}{{N}_i}

\newcommand{\ntxants}{{M}}
\newcommand{\totpow}{P_T}
\newcommand{\totUsers}{N}

\newcommand{\ngrps}{G}

\makeatletter
\def\endthebibliography{%
	\def\@noitemerr{\@latex@warning{Empty `thebibliography' environment}}%
	\endlist
}
\makeatother

\begin{document}
\allowdisplaybreaks

\title{Joint User Grouping, Scheduling, and Precoding for Multicast Energy Efficiency in Multigroup Multicast Systems}
 \author{$\text{Ashok Bandi}^{}$, $\text{Bhavani Shankar Mysore R, }^{}$ $\text{Symeon Chatzinotas}^{}$ $\text{and Bj{\"o}rn Ottersten}^{}$\\        
        Interdisciplinary Centre for Security, Reliability and Trust (SnT), the University of Luxembourg, Luxembourg.
        \\ Email: \{ashok.bandi, bhavani.shankar, symeon.chatzinotas, bjorn.ottersten\}@uni.lu
        \thanks{This work is supported in part by Luxembourg national fund FNR project PROSAT.
    }\thanks{This work has been published in part at IEEE Global communications conference (GLOBECOM), 2019 \cite{FrameMGMC_Ash_globecom19} and IEEE conference on signal processing and communication (SPCOM) 2020 \cite{MSE_SPCOM_Ashok}.
    }
        }


	\maketitle
 \begin{abstract}          
      This paper studies the joint design of user grouping, scheduling (or admission control) and precoding to optimize energy efficiency (EE) for multigroup multicast scenarios in single-cell multiuser MISO downlink channels. Noticing that the existing definition of EE fails to account for group sizes, a new metric called multicast energy efficiency (MEE) is proposed. In this context, the joint design is considered for the maximization of MEE, EE, and scheduled users. Firstly, with the help of binary variables (associated with grouping and scheduling) the joint design problem is formulated as a mixed-Boolean fractional programming problem such that it facilitates the joint update of grouping, scheduling and precoding variables. Further, several novel optimization formulations are proposed to reveal the hidden difference of convex/ concave structure in the objective and associated constraints. Thereafter, we propose a convex-concave procedure framework based iterative algorithm for each optimization criteria where grouping, scheduling, and precoding variables are updated jointly in each iteration. Finally, we compare the performance of the three design criteria concerning three performance metrics namely MEE, EE, and scheduled users through Monte-Carlo simulations. These simulations establish the need for MEE and the improvement from the system optimization.
    \end{abstract}
    

\section{Introduction}

The mobile data traffic is exploding unprecedentedly due to the exponential increase in mobile devices and their demand for throughput hungry service/ applications \cite{BillDev}. This has led to the adaptation of full-spectrum reuse and multi-antenna technologies, which result in significantly improved spectral efficiency (SE). On the other hand, the demand for green communications necessitates achieving these high SEs with limited energy \cite{GreenCom}. In this regard, energy efficiency (EE), which measures the performance in throughput/Watts, becomes a key factor to be considered in the next-generation wireless networks \cite{EE_Oskari}. Notice that the power minimization or energy minimization is also referred to as energy efficiency in the literature. The aforementioned EE which measure the performance in throughput/Watts is the focus of this paper.

On the other hand, in some scenarios like live-streaming of popular events, multiple users are interested in the same data. Realizing that multicasting such information to groups of users leads to better utilization of the resources, physical layer multigroup multicasting (MGMC) has been proposed in \cite{MGMC_NPhard,8359237}. Noticing the significant improvement in EE, multicasting has been adopted into 3GPP standards \cite{6353684}. However, the following challenges need to be addressed to fully leverage the gains of MGMC:
\begin{itemize}
    \item Inter-group interference: The co-channel users in different groups generate interference across the groups which is referred to as inter-group interference (IGI). A study of IGI is essential as it fundamentally limits the minimum rate of the groups that can be achieved \cite{MGMC_feasibility} and, hence, the total throughput of the network. In this context, \emph{user grouping} is a pivotal factor to be considered since it dominantly influences IGI \cite{MGMC_JSP_Dimi2015,Hu2018}. 
    \item Infeasibility: In a real scenario, each user needs to be served with a certain quality-of-service (QoS); failing to meet the QoS leads to retransmissions which significantly decrease the EE of the network. The severe IGI and/ or poor channel gains may thwart some users from meeting their QoS \cite{MGMC_feasibility}. On the contrary, even in the cases with lower IGI, limited power may restrain the users from meeting their QoS \cite{MGMC_NPhard}. Due to a combination of these three factors, the system may fail to satisfy the QoS requirements of all the users in all groups. This scenario is referred to as the infeasibility of the MGMC design in the literature \cite{MGMC_feasibility, MGMC_NPhard}. The infeasibility of MGMC is crucial to the design and is, therefore, typically addressed by \emph{user scheduling} (also referred to as admission control in the literature) \cite{Maskani_SchBF_MGMC, Hu2018}.
    \end{itemize}

\subsection{Joint user grouping, group scheduling and user scheduling for message-based MGMC systems}
In this work, similar to \cite{MessageMGMC_mixiaTao, Hu2018}, a message based user grouping and scheduling are considered. In the message based MGMC model, each group is associated with a different message, and a user may be interested in multiple messages, thus requiring user grouping. Unlike \cite{MessageMGMC_mixiaTao} and similar to \cite{Hu2018}, a limited antenna BS system is considered with the number of groups being larger than the number of antennas at BS. Therefore, in a given transmit slot, only a few groups (equal to the number of antennas of BS) are scheduled; this is referred to as group scheduling. Further, users that fail to satisfy the QoS requirements of all the interested groups are simply excluded from the grouping; this is referred to as user scheduling. So, the considered model requires the design of user grouping, group scheduling, and user scheduling. User grouping, group scheduling, and user scheduling are inter-related. To see this, user grouping decides the achievable minimum signal-to-interference and noise ratio (SINR) of the groups (or IGI) which influences the group scheduling and user scheduling. Further, omitting or adding a user (i.e., user scheduling) in a group changes the IGI, thereby impacting group scheduling. Similarly, user scheduling in a group might necessitate the re-grouping of users (i.e., user grouping); this affects IGI and group scheduling. Furthermore, IGI is a function of precoding \cite{MGMC_JSP_Dimi2015}. Therefore, the optimal performance requires the \emph{joint design of user grouping, group scheduling, user scheduling, and precoding; this is compactly referred to as joint design in this paper}. 


\subsection{EE in the context of the joint design of user grouping, scheduling, and precoding}
In this work, for the reasons mentioned earlier, EE is considered as the measure of the system performance. All the existing works on EE maximization in MGMC systems \cite{8359237, EE_MGMC_frist, EE_MGMC_StatCSI, EE_Oskari, LDM_EE_MGMC}, presume a particular user grouping and scheduling, therefore, the EE for MGMC systems is defined as the ratio of the sum of minimum throughput within each group and the total consumed energy. Notice that the existing EE definition accounts for only the minimum rate of a group ignoring the number of users in the groups (group sizes). However, in the context of user grouping and scheduling, group sizes need to be accounted for, in addition to their minimum rates. To comprehend the necessity, consider two groups with an equal minimum rate and a large difference in group sizes. According to the existing EE definition, the group with few users could be scheduled as the EE maximization is not biased to schedule a group with the larger size. However, from the network operator perspective, scheduling the group with more users results in efficient utilization of the resources. Moreover, the event of scheduling the group with few users is likely for EE maximization since it usually consumes less energy. However, if a large number of users can be served with a slight increase in energy, scheduling such a larger group improves the efficiency in the utilization of resources. The existing frameworks can not handle these scenarios as the number of users is not included in the EE definition. Noticing the drawbacks of existing EE definition, in this work, a new metric called multicast energy efficiency (MEE) is proposed to account for the group sizes along with the minimum rates. \textcolor{blue}{In contrast to EE, in the numerator of MEE minimum throughput within a group times number of users in that group, is considered. Realizing the importance of MEE, in this work, we consider the joint design of user grouping, scheduling, and precoding for MGMC systems subject to grouping, scheduling, quality-of-service (QoS) and total power constraints for the maximization of three design criteria: MEE, EE and scheduled users}. In this context, related works in the literature and contributions/novelty of the paper are summarized in the sequel of this section.

\subsection{Related works}\label{sec:RelWork}

\subsubsection*{Energy efficiency for MGMC systems}
\textcolor{blue}{The EE maximization problem for MGMC systems was first addressed in the context of coordinated beamforming for multicell networks\cite{EE_MGMC_frist}. By definition, EE belongs to fractional programming. The authors in \cite{EE_MGMC_frist} used Dinkelbach's method to transform this fractional program to subtractive non-linear form and further solved the problem with an iterative algorithm wherein each iterate precoding and power vectors are updated alternatively. Later, this work is extended to the case of imperfect channel state information in \cite{EE_MGMC_StatCSI}. Dinkelbach's method based transformation works efficiently if the denominator is a simple linear objective and numerator is convex otherwise it leads in a multi-level parametric iterative algorithm that is not efficient \cite{RateDepPowModel_MGMC}. In \cite{RateDepPowModel_MGMC}, the authors optimized the EE for MGMC in multicell networks considering the rate-dependent processing power. The authors in \cite{RateDepPowModel_MGMC} use successive convex approximation to transform fractional EE maximization problem as convex-concave programming in iteration $k$ and further solved the subproblem using Charnes-Cooper transformation (CCT). In \cite{EE_Oskari}, EE maximization in a large antenna system with antenna selection is solved using SCA based CCT. Further, the authors in \cite{EE_Oskari} addressed the boolean nature stemming from antenna selection by continuous relaxation and followed by thresholding. In \cite{LDM_EE_MGMC}, EE maximization for non-orthogonal layered-division multiplexing based joint multicast and the unicast system is considered. A pseudoconvex approach based parallel solution is developed for EE maximization in MIMO interference channels in \cite{Yang_EE}. EE user scheduling and power control is considered for multi-cell OFDMA networks for a unicast scenario in \cite{8471203}. Moreover, precoding is not considered in \cite{8471203}. The methodologies used in all these works are either SCA based CCT \cite{EE_MGMC_frist,EE_Oskari,RateDepPowModel_MGMC} or  SCA based Dinkelbach's method \cite{LDM_EE_MGMC,7328326, Yang_EE}. Unlike the aforementioned works which assume the rate-dependent processing power to be a convex function of rate, a non-convex power consumption model is considered in this work. Therefore, unlike EE maximization considered in the literature, MEE maximization considered in this work belongs to mixed-integer fractional programming where the numerator is mixed-integer non-convex and denominator is also non-convex. Hence, the SCA based CCT can not be applied and Dinkelbach's method yields parametric multilevel iterative algorithms\cite{Yang_EE}. Moreover, the integer nature stemming from the user grouping and scheduling is different to the antenna selection problems (see \cite{EE_Oskari} and references therein) and, hence, problem formulation and solution methodologies used in the antenna selection literature can not be employed.}

\subsubsection*{User grouping, scheduling, and precoding for EE maximization in MGMC systems}
In this work, we consider the joint design of user grouping, group scheduling, user scheduling, and precoding for MEE maximization in message-based MGMC systems. Joint design of admission control and beamforming for MGMC systems was initially addressed in \cite{Maskani_SchBF_MGMC} for the power minimization problem. The authors in \cite{Maskani_SchBF_MGMC} addressed the admission control using binary variables and transformed the resulting mixed-integer non-linear problem (MINLP) into a convex problem using semidefinite programming (SDP) transformations; SDR based precoding is likely to include high-rank matrices for MGMC systems \cite{Kaliszan_2012}; hence, the solutions may become infeasible to the original problems.  Later, the design of user grouping and precoding without admission control is considered in \cite{MGMC_JSP_Dimi2015} for satellite systems. However, the authors in \cite{MGMC_JSP_Dimi2015} adopted the decoupled approach of heuristic user grouping followed by a semidefinite relaxation (SDR) based precoding. In \cite{MessageMGMC_mixiaTao}, the authors considered the joint user grouping and beamforming without user scheduling for massive multiple-input multiple-output (MIMO) systems and proved that arbitrary user grouping is asymptotically optimal for max-min fairness criteria. However, the arbitrary grouping is not optimal for other design criteria and also not optimal for max-min criteria in the finite BS antenna system. Moreover, the underlying precoding problem in \cite{Maskani_SchBF_MGMC,MGMC_JSP_Dimi2015,Hu2018} is solved by SDR transformation, hence, as mentioned earlier the solutions may become infeasible to the original problems \cite{Kaliszan_2012}. In \cite{9013133}, joint adaptive user grouping and beamforming is considered for MGMC scenario in massive MIMO system. The authors in \cite{9013133} adapted iterative approach wherein each iterate user grouping and beamforming are solved separately by decoupling the two problems. However, the EE or MEE maximization problem in the context of user grouping, scheduling, and precoding is not considered in the literature.
 
\textcolor{blue}{ The system model considered in this work is similar to \cite{Hu2018}. The authors in \cite{Hu2018} considered the joint design for power minimization and its extension to EE maximization is not clear. Moreover, at the solution level, the problem is decoupled into user grouping and scheduling followed by SDR based precoding which is likely to include high-rank matrices for MGMC systems \cite{Kaliszan_2012}; hence, the solutions may become infeasible to the original problems. 
In our previous work \cite{AshJSP_MISO_2019}, we considered the joint design of scheduling and precoding for the unicast scenario to optimize sum rate, Max-min SINR, and network power. In \cite{AshJSP_MISO_2019}, scheduling is addressed by bounding the power of the precoder with the help of a binary variable. However, in the MGMC system, each precoder is associated with a group of users, hence, the same method can not be employed. Moreover, the MEE or EE maximization in the context of user grouping, scheduling, and precoding belongs to mixed-integer fractional programming which is not dealt in \cite{AshJSP_MISO_2019}, and its extension to proposed system model is not clear.  Furthermore, the proposed MEE belongs to Mixed-integer fractional programming problems with a mixed-integer non-convex objective in the numerator and a non-convex objective in the denominator. Therefore, the MEE maximization problem considered in this work is significantly different from \cite{AshJSP_MISO_2019} in terms of the system model, performance metric of optimization, problem formulation, and the nature of the optimization problem. Hence, the solution in \cite{AshJSP_MISO_2019} can not be applied directly here.
The MEE maximization problem considered for the joint design in this work is highly complex as it inherits the complications of user grouping, group scheduling and user scheduling, and EE problems, and poses additional challenges.} 


\subsection{Contributions}\label{sec:Contrb}
Below we summarize the contribution on the joint design of user grouping, scheduling, and precoding for the MGMC system to maximize the MEE and EE  as follows:
\begin{itemize}
\item \textcolor{blue}{Noticing that the existing EE definition accounts only for the minimum rate of groups ignoring group size, in the context of user grouping and scheduling a new metric called MEE is proposed to account for the group sizes along with the minimum rate of the groups in the messaged-based MGMC systems \cite{Hu2018}. Unlike the existing works e.g., \cite{Hu2018, RateDepPowModel_MGMC, AshJSP_MISO_2019}, this results in a new mixed-Boolean fractional objective function posing additional challenges to the existing challenges in user grouping, scheduling, and EE designs. }
\item \textcolor{blue}{Further, unlike existing models which assumes rate-dependent processing power to be a convex function of rate \cite{EE_Oskari,EE_MGMC_frist,8471203,LDM_EE_MGMC,RateDepPowModel_MGMC}, rate-dependent processing power is assumed to be a non-convex function of rate with admissible  DC decomposition. Therefore, the considered power consumption model applies to a broader class of models.}
\item \textcolor{blue}{Inspired by the work in \cite{AshJSP_MISO_2019}, user grouping, group scheduling, and user scheduling are addressed with the help of binary variables. Unlike \cite{AshJSP_MISO_2019}, MEE maximization problems along with binary constraints result in a new mixed-Boolean fractional programming to which the existing SCA based CCT \cite{RateDepPowModel_MGMC} can not be applied and Dinkelbach's \cite{EE_MGMC_frist} method results in the parametric multilevel iterative algorithm which is not efficient.}

\item The resulting mixed-Boolean fractional formulations are non-convex and NP-hard. Towards obtaining a low-complexity stationary solution, with the help of novel reformulations, the fractional and non-convex nature of the problems is transformed as DC functions. Further, Boolean nature is handled with appropriate relaxation and penalization. These reformulations render the joint design as a DC problem, a fact hitherto not considered.
\item Finally, within the framework of the convex-concave procedure (CCP) \cite{Yuille_CCCP_2001} (which is a special case of SCA \cite{SCA_org}), an iterative algorithm is proposed to solve the resulting DC problem wherein each iterate a convex problem is solved. A simple low-complexity non-iterative procedure to obtain a feasible initial point, which inherently establishes convergence of the proposed algorithms to a stationary point \cite{Sriperumbudur_CCP_2009}, is proposed.
\item The performance of the proposed algorithms affecting the three design aspects, namely MEE, EE, and number of scheduled users, and their typical quick convergence behavior (which confirms the low-complexity nature) are numerically evaluated through Monte-Carlo simulations. 
\end{itemize}



The sequel is organized as follows. Section~\ref{sec:system_model} presents MGMC system. Further, the joint design for MEE problem in Section~\ref{sec:MEE}, EE problem in Section~\ref{sec:EE} and SUM problem in Section~\ref{sec:SUM}. Section~\ref{sec:simulation} presents simulations and  Section~\ref{sec:conclusion} concludes the work.

\emph{Notations:} Lower or upper case letters represent scalars, lower case boldface letters represent vectors, and upper case boldface letters represent matrices. $\|\cdot\|$ represents the Euclidean norm, $|\cdot|$ represents the cardinality of a set or the magnitude of a scalar, $(\cdot)^H$ represents Hermitian transpose, $(\cdot)^T$ represents  transpose, $\binom{a}{b}$ represents  $a$ choose $b$, and $\Real\lbrace \rbrace$ represents real operation, $\mathbb{E}\lbrace \rbrace$ represents expectation operator and $\nabla$ represents the gradient.
\section{System model}\label{sec:system_model}
\subsection{Message based user grouping and scheduling}
We consider the downlink scenario of a single cell multiuser MISO system with $\ntxants$ transmit base station (BS) antennas and $\totUsers \left(\geq \ntxants\right)$ users each equipped with a single receive antenna.  In this work, similar to \cite{MessageMGMC_mixiaTao, Hu2018}, message-based user grouping, and scheduling is considered. \textcolor{blue}{In this context, it is assumed that each group is associated with a unique message. Therefore, the number of groups, say $\ngrps$, is equal to the number of messages.  Further, each user is assumed to be interested in at least one message and a user may be interested in multiple messages. Despite the user's interest in multiple messages, a user is allowed to be a member of the utmost one group.} This constraint is simply referred to as user grouping constraint (UGC). Letting $\cal{S}_i$ to be the set of users belonging to message (group) $i$ and $\phi$ to be the empty set, the UGC is formulated as $\cal{S}_i \cap \cal{S}_j=\phi, \text{ for } i \neq j $. Further, to establish the relevance of the design to the real scenarios, a certain QoS requirement (typically depending on the type of service/application) on the messages is assumed. UGC also captures the worst-case scenario of a user failing to meet any QoS requirement associated with any of the interesting messages: hence, the user is simply not scheduled in the current slot. Therefore, UGC naturally leads to $\sum_{i=1}^{\ngrps}\card{\cal{S}_i} \leq \totUsers$. Further, it is assumed that  $\ngrps \geq \ntxants$, hence, scheduling of exactly $\ntxants$ groups out of $\ngrps$ is considered. This constraint is simply referred to as group scheduling constraint (GSC). 

User channels are assumed to constant and perfectly known. The noise at all users is assumed to be independent and characterized as additive white complex Gaussian with zero mean and variance $\sigma^2$. Furthermore, total transmit power at the BS is limited to $\totpow$ for each transmission. Finally, the BS is assumed to transmit independent data to different groups with $\mathbb{E}\lbrace|x_i|^2\rbrace=1,\text{}\forall i$, where $x_i$ is the message associated with group $i$. Let $\mathbf{w}_{i} \in \mathbb{C}^{\ntxants \times 1}$ be the precoding vector with group $i$ and $\mathbf{W}=\left[\mathbf{w}_{1},\hdots,\mathbf{w}_{\ngrps}\right]$, $\mathbf{h}_{i} \in \mathbb{C}^{\ntxants \times 1}$ be the downlink channel of user $i$, and  $\gamma_{ij}=\dfrac{|\mathbf{h}_{i}^H\mathbf{w}_j|^2}{\sum_{l\neq j}|\mathbf{h}_{i}^H\mathbf{w}_l|^2+\sigma^2}$ be the SINR of user $i$ belonging to group $j$.

\subsection{Power consumption model, Energy efficiency, and Multicast energy efficiency}\label{sec:EEE_def}
\subsubsection*{Power consumption model}
\textcolor{blue}{In this work, we adopt the power consumption model proposed in \cite{RateDepPowModel_MGMC}. Let $\BW$ be the bandwidth of the channel and ${r}_j= \BW \log_2\left(1+\min_{i \in \mathcal{S}_j}\gamma_{ij}\right)$ be the minimum rate of group $j$. Notice that all the users in a group receive exactly the same message associated with the group. Therefore, the transmission rate of the message $j$ to group $j$ at the BS is simply the minimum rate of the group i.e., $r_j$.  With defined notations, the power consumption at the BS is defined as
\begin{equation}
    {g}(\mathbf{W},\boldsymbol{r}) \triangleq P_0 + \sum_{j=1}^{\ngrps} \left(\dfrac{1}{\rho} \norm{\mathbf{w}_i}^2 + \Pi {p}\left(r_j\right)\right), 
\end{equation}
where $\boldsymbol{r}=\left[r_1,\hdots,r_{\ngrps}\right]$, $P_0$ is the static power spent by the cooling systems, power supplies etc., $\rho <1$ is the power amplifier efficiency, and $\Pi \geq 0$ is a constant accounting for coding and decoding power loss, and ${p}(r_j)={p}_{1}\left(r_j\right)-{p}_{2}\left(r_j\right)$ is a differentiable non-negative difference-of-convex function of $r_j$  reflecting the rate-dependent processing power of group $j$ with  ${p}(0)=0$, and  ${p}_{1}$ and ${p}_{2}$ are convex functions. Notice that unlike previous works e.g. \cite{RateDepPowModel_MGMC,8471203,EE_Oskari} where ${p}(r_j)$ is assumed to be a convex functions, the considered model ${g}(\mathbf{W},\boldsymbol{r})$ represents relatively broader class of rate-dependant power consumption models.}
\subsubsection*{Energy efficiency}
EE for MGMC systems is typically defined as a ratio of the throughput of the network  to the energy consumed at the BS in the literature. Letting $\mathcal{T}$ to be the set of scheduled groups and $\BW$ be the bandwidth of the channel, the EE is defined as 
\eqref{eq:EE_def}.
 \begin{equation}
      \text{EE}\triangleq \dfrac{ \sum_{j \in \mathcal{T}} \text{} \BW \log\left(1+  \min_{i \in \mathcal{S}_j}\gamma_{ij}\right) }{{g}(\mathbf{W},\boldsymbol{r})  }.\label{eq:EE_def}
\end{equation}
The numerator of the EE in \eqref{eq:EE_def} models the network's multicast throughput as the sum of the minimum throughput of all groups. So, this definition only accounts for the minimum throughput of a group ignoring its size. 
\subsubsection*{Multicast energy efficiency }
In the context of user grouping and scheduling for MGMC systems, the standard EE metric needs to be redefined to account for the size of the group. To understand this, consider a scenario of scheduling a group between two groups having the same minimum throughput, consuming the same energy and large difference in group sizes. The EE criterion does not discriminate between two groups. However, scheduling a group with a large number of users leads to better utilization of resources. So, to account for the number of users being served in each group along with its minimum rate, we propose a new metric called MEE for the MGMC systems. With the help of defined notations, MEE is formally defined as,
 \begin{equation}
      \text{MEE}\triangleq \dfrac{ \sum_{j \in \mathcal{T}} \left( \Psi_j \card{\mathcal{S}_j} \text{} \BW \log\left(1+  \min_{i \in \mathcal{S}_j}\gamma_{ij}\right) \right)}{{g}(\mathbf{W},\boldsymbol{r})  }.\label{eq:EEE_def}
\end{equation}
\textcolor{blue}{where $\Psi_j$ is the weight associated with group $j$. The weights i.e., $\Psi_j$s are introduced in MEE to address the fairness among the groups. For example, by choosing $\Psi_1$ to be relatively much larger than $\lbrace \Psi_i \rbrace_{j=2}^{\ngrps}$ scheduling of group 1 can be prioritized.}
\subsubsection*{Interpretation of MEE as total received bits/Joule} 
From the physical layer transmission perspective, the network throughput (number of transmitted bits per second) in MGMC systems is same as unicast systems. In unicast scenario, the transmitted information is received by only one user. However, in MGMC scenario, the information transmitted to group $j$ is received by $\card{\mathcal{S}_j}$ users. Hence, from the network operator perspective, throughput of group $j$ in this multicast scenario is $\card{\mathcal{S}_j} \BW \log\left(1+  \min_{i \in \mathcal{S}_j}\gamma_{ij}\right)$ received bits per second. Motivated by this, the numerator of equation \eqref{eq:EEE_def} i.e., $\sum_{j \in \mathcal{T}} \left( \card{\mathcal{S}_j} \text{} \BW \log\left(1+  \min_{i \in \mathcal{S}_j}\gamma_{ij}\right) \right)$ reflects the combined multicast throughput of all the groups i.e., network throughput, henceforth, simply referred to in this work as multicast throughput. Similarly, MEE defined in \eqref{eq:EEE_def} reflects MEE for MGMC systems. Thus, MEE can be seen as number of received bits for one joule of transmitted energy.

\section{Multicast Energy Efficiency}\label{sec:MEE}

In this section, at first, the joint design of user grouping, scheduling, and precoding is mathematically formulated to maximize the MEE subject to appropriate constraints on the number of groups, users per group, number of scheduled groups, power, and QoS constraints. This problem is simply referred to as the MEE problem. Further, with the help of useful relaxations and reformulations, the MINLP NP-hard MEE problem is transformed as a DC programming problem. Finally, within the framework of CCP, an iterative algorithm is proposed which guarantees to attain a stationary point of the original problem.    

\subsection{Problem formulation: MEE}
The EE maximization problem, with the notations defined, in the context of user grouping, scheduling and precoding for the MGMC scenario in Section~\ref{sec:system_model} is formulated as, 
\begin{align}\label{eq:EEE_Prob_def}
  \mathcal{P}_{1}^{\text{MEE}}:&\max_{\lbrace\mathbf{w}_j, \mathcal{S}_j \rbrace_{j=1}^{\ngrps}} \dfrac{ \sum_{j=1}^{\ngrps} \left(\Psi_j \card{\mathcal{S}_j} \text{} \BW \log\left(1+  \min_{i \in \mathcal{S}_j}\gamma_{ij}\right) \right)}{\color{blue}{g}(\mathbf{W},\boldsymbol{r}) } \\
    \text{ s.t. } 
    C_1:\text{ }&\mathcal{S}_i \cap \mathcal{S}_j = \phi, i\neq j, \forall i, \text{ } \forall j, \nonumber \\
    C_2:\text{ }& \norm{\left[\card{\mathcal{S}_1}, \hdots, \card{\mathcal{S}_\ngrps}\right]}_0 = \ntxants, \nonumber \\
    C_3:\text{ } &\log\left(1+\gamma_{ij}\right) \geq \epsilon_j, \text{ } i \in \mathcal{S}_j, \text{ } \forall j, \nonumber  \\
    C_4:\text{ }&\sum_{j =1}^{\ngrps}\norm{\mathbf{w}_{j}}^2 \leq \totpow,\nonumber \\
    \color{blue}C_5:\text{ }& \color{blue}\mathcal{S}_i \subset \lbrace1,\hdots,\totUsers \rbrace, \text{} \forall i, \nonumber 
     \end{align} 
    where $\epsilon_j$ is the QoS requirement of group $j$, $\forall i$ refers to $i \in \lbrace1,\hdots,\totUsers\rbrace$ and $\forall j$ refers to $j \in \lbrace1,\hdots,\ngrps\rbrace$
    \\
    \emph{Remarks}:
\begin{itemize}
    \item Constraint $C_1$ is the UGC; constrains a user to be a member of at most one group.
    \item Constraint $C_2$ is the GSC; it ensures the design to schedule exactly $\ntxants$ groups.  
    \item Constraint $C_3$ is the QoS constraint; it enforces the scheduled users in each group to satisfy the corresponding minimum rate requirement associated with the group. This enables the flexibility to support different rates on different groups. Hereafter, the constraint $C_3$ is simply referred to as QoS constraint. \textcolor{blue}{Moreover, the constraint $C_3$ together with $C_1$ ensure the USC.}
    \item Constraint $C_4$ is the total power constraint (TPC); precludes the design from consuming the power in excess of available power i.e., $\totpow$. 
\end{itemize}

\paragraph*{Necessity of low-complexity algorithms for joint design} The problem $\mathcal{P}_{1}^{\text{MEE}}$ is combinatorial due to constraints $C_1$ and $C_2$. Hence, obtaining the optimal solution to $\mathcal{P}_{1}^{\text{MEE}}$ requires an exhaustive search-based user grouping and scheduling. To understand the complexity of the exhaustive search methods, assume that each user is interested in only one message. Further, let $\totUsers_i$ be the number of users in the group $i$.  Let $\mathcal{T}_i$ be the all possible scheduling subsets of $\mathcal{S}_i$, so the number of sets in $\mathcal{T}_i$ is $\sum_{j=0}^{\UesPerGrp}\binom{\UesPerGrp}{j}$ for $i\in\lbrace 1,\hdots,\ngrps \rbrace$. So, the exhaustive search needs to be performed over the Cartesian product of sets $\mathcal{T}_i$s i.e., $\times_{i=1}^{\ngrps}\left( \mathcal{T}_i\right)$. It is easy to see that the exhaustive search algorithms quickly become impractical due to exponential complexity. This case merely a simple case of the problem considered in $\mathcal{P}_{1}^{\text{MEE}}$. Additionally, for each scheduled combination, the corresponding precoding problems in $\mathcal{P}_{1}^{\text{MEE}}$ need to be solved. Moreover, these precoding problems are generally not only NP-hard but also non-convex \cite{MGMC_NPhard}. Thus, in the sequel, we focus on developing low-complexity algorithms that are guaranteed to obtain a stationary point of the NP-hard and non-convex problem $\mathcal{P}_{1}^{\text{MEE}}$.


\subsection{A mixed integer difference of concave formulation: MEE}\label{sec:DCform_EE}

In this section, firstly, avoiding the set notation by using binary variables the problem  $\mathcal{P}_{1}^{\text{MEE}}$ is equivalently reformulated as an MINLP problem without the set notations. Further, with the help of a minimal number of slack variables and novel reformulations, the resulting MINLP problem is transformed as a difference-of-concave (DC) problem subject to binary constraints.

Towards transforming the MEE problem in $\mathcal{P}_{1}^{\text{MEE}}$ as DC a problem, let $\eta_{ij}$ be the binary variable indicating the membership of user $i \in \lbrace 1, \hdots,\totUsers \rbrace$ in group $j\in\lbrace1,\hdots,\ngrps\rbrace$. In other words, $\eta_{ij}=1$ indicates that user $i$ is a member of the group $j$ and not a member otherwise. Since a user may not be interested in some groups, the $\eta_{ij}$s corresponding to these groups is fixed beforehand to zero. Hence, only a subset of the entries in $\boldsymbol{\eta}_i$ are the variables of the optimization. However, for the ease of notation, without the loss of generality, henceforth, we assume that each user is interested in all the groups. In other words, all the entries in $\boldsymbol{\eta}_i$ become variables of optimization. It is easy to see that this is only a generalization to the aforementioned case. Hence, a solution to this generalized problem is a solution to the aforementioned problem. 

Letting $\Theta_j$ and $\zeta_j$ be the slack variables associated with minimum rate of group $j$ respectively, and $\alpha_{ij}$ be the slack variable associated with SINR of user $i$ of group $j$, the problem $\mathcal{P}_{1}^{\text{MEE}}$ is equivalently reformulated as,
 \begin{align}\label{eq:EEE_MINLP1}
  \mathcal{P}_{2}^{\text{MEE}}:\text{} &\max_{\mathbf{W},\boldsymbol{\Theta,\eta,\alpha, \zeta} } \text{ } \dfrac{ \sum_{j=1}^{\ntxants} \left(\sum_{i=1}^{\UesPerGrp}\eta_{ij}\right)\BW \Psi_j\Theta_j }{\color{blue}{g}(\mathbf{W},\boldsymbol{\zeta}) } \\
    \text{ s.t. } 
    &C_1:\text{} \eta_{ij} \in \lbrace 0,1 \rbrace, \text{ } \forall i, \text{ }\forall j, 
    &&C_2: \text{} \sum_{j=1}^{\ngrps}\eta_{ij} \leq 1,\forall i, \nonumber \\
    &C_3: \text{} \norm{\left[\sum_{i=1}^{\totUsers}\eta_{i1},\hdots,\sum_{i=1}^{\totUsers}\eta_{i\ngrps}\right]}_0 = \ntxants, 
    &&C_4: \text{}1+\gamma_{ij} \geq \alpha_{ij}, \text{ } \forall i, \text{ } \forall j, \nonumber \\
    &C_5: \text{}\log\alpha_{ij} \geq \eta_{ij} \Theta_{j}, \text{ } \forall i, \text{ } \forall j, 
    &&C_6: \text{}\Theta_j \geq \epsilon_j, \text{ } \forall j, \nonumber \\
    &C_7:\text{}\sum_{i=1}^{\ntxants}\norm{\mathbf{w}_i}_2^2 \leq \totpow,
    &&\color{blue}C_8:\text{ } 0 \leq \zeta_j \leq \norm{\sum_{i=1}^{\totUsers}\eta_{ij}}_0 \BW\Theta_j, \forall j, \nonumber
    \end{align} 
    where $\boldsymbol{\Theta}=\left[\Theta_1,\hdots,\Theta_{\ntxants} \right]$, $\boldsymbol{\zeta}=\left[\zeta_1,\hdots,\zeta_{\ntxants} \right]$, $\boldsymbol{\eta}=\left[\boldsymbol{\eta}_1,\hdots,\boldsymbol{\eta}_{\ngrps} \right]$, $\boldsymbol{\eta}_i=\left[ {\eta}_{i1},\hdots,{\eta}_{i\ngrps} \right]$, $\boldsymbol{\alpha}=\left[\boldsymbol{\alpha}_1,\hdots,\boldsymbol{\alpha}_{\ngrps} \right]$, and $\boldsymbol{\alpha}_i=\left[ {\alpha}_{i1},\hdots,{\alpha}_{i\ngrps} \right]$ \\
        \emph{Remarks:}
    \begin{itemize}
    \item Constraints $C_1$ and $C_2$ in $\mathcal{P}_{2}^{\text{MEE}}$ ensures the UGC. The constraint $C_3$ is the equivalent reformulation of GSC constraint $C_2$ in $\mathcal{P}_{1}^{\text{MEE}}$.
    \item For all the users that are not subscribed to group $j$ ( i.e., users with $\eta_{ij}=0$), the constraint $C_5$ implies $\log \alpha_{ij}\geq 0$ which is satisfied by the definition of rate. On the contrary, for all the users subscribed to group $j$ (i.e., users with $\eta_{ij}=1$) constraint $C_4$ implies $\log \alpha_{ij} \geq \Theta_j$. Hence, $\Theta_j$ provides the lower bound for the minimum rate of the group. Moreover, at the optimal solution of $\mathcal{P}_{2}^{\text{MEE}}$, $\Theta_j$ is equal to the minimum rate of group $j$ i.e., $\Theta_j=\min_{j\in \mathcal{S}_i}  \log \alpha_{ij}$.
     \item In the objective of $\mathcal{P}_{2}^{\text{MEE}}$, the term $\sum_{i=1}^{\UesPerGrp}\eta_{ij}$ is equivalent to $\card{\mathcal{S}_i}$.  Since at the optimal solution $\Theta_j=\min_{j\in \mathcal{S}_i} \log \alpha_{ij}$, the objective in $\mathcal{P}_{2}^{\text{MEE}}$ is equivalent to the EE objective in $\mathcal{P}_{1}^{\text{MEE}}$.
     \item \textcolor{blue}{ Constraint $C_8$ is introduced to address the rate-dependent processing power in $g\left(\mathbf{W},\boldsymbol{r}\right)$ in problem $\mathcal{P}_1^{\text{MEE}}$. For a unscheduled group $j$ (i.e., $\norm{\sum_{i=1}^{\totUsers}\eta_{ij}}_0=0$), from constraint $C_8$ $\zeta_j=0$ and for a scheduled group i.e., ($\norm{\sum_{i=1}^{\totUsers}\eta_{ij}}_0=1$) $\zeta_j=\Theta_j$ which is the minimum rate of the group.}
     \end{itemize}
    
    \textcolor{blue}{Notice that the problem $\mathcal{P}_2^{\text{MEE}}$ is significantly different and much more complex than problems dealt in \cite{EE_Oskari,EE_MGMC_frist,RateDepPowModel_MGMC,Yang_EE,AshJSP_MISO_2019,Hu2018,MessageMGMC_mixiaTao}. The  MEE objective in $\mathcal{P}_2$ is unlike any EE objective in the literature (see \cite{EE_Oskari,EE_MGMC_frist,RateDepPowModel_MGMC,Yang_EE} and reference therein). The power consumption model $g\left(\mathbf{W},\boldsymbol{r}\right)$ and multicast throughput i.e., $ \sum_{j=1}^{\ntxants} \left(\sum_{i=1}^{\UesPerGrp}\eta_{ij}\right)\BW \Theta_j$ considered in this work are non-convex and multicast throughput is a function of binary variables. Hence, SCA based CCT \cite{EE_Oskari} can not apllied $\mathcal{P}_2$ and Dinkelbach's methods \cite{EE_MGMC_frist} results in a parametric multi level iterative algorithm. Further, problem $\mathcal{P}_2$ differs from \cite{AshJSP_MISO_2019} where the binary variables are only associated with precoding and SINR terms. The transformation to deal with the MEE objective, constraint $C_3$ and $C_8$ are not dealt in \cite{AshJSP_MISO_2019}. The problem $\mathcal{P}_2^{\text{MEE}}$ inhibits the complexities associated with EE problems and user grouping and scheduling problems, hence, much more challenging than standalone EE and user grouping and scheduling problems.}
    
 The reformulation given in $\mathcal{P}_2^{\text{MEE}}$ is equivalent to $\mathcal{P}_1^{\text{MEE}}$ that the optimal solution of $\mathcal{P}_2$ is also the optimal solution of $\mathcal{P}_1$. Hence, the problem  $\mathcal{P}_2$ is an equivalent reformulation of $\mathcal{P}_1$. The problem $\mathcal{P}_2$ is combinatorial due to constraint $C_1$ and $C_3$, and non-convex due to constraint $C_3$, $C_4$ and the objective. Letting $\delta_{i}$ to be the slack variable associated with group $i$ and $t$ to be the slack variable associated with power consumption, $\mathcal{P}_2$ is transformed into a DC problem subject to binary constraints as,
      \begin{align}\label{eq:EEE_MIDC}
    &\mathcal{P}_{3}^{\text{MEE}}: \max_{\mathbf{W},\boldsymbol{\Theta,\eta,\delta,\alpha, \zeta} } \text{ }  \sum_{i=1}^{\totUsers}\sum_{j=1}^{\ngrps}f\left(\eta_{ij},\Theta_j,t\right) \triangleq 
     \BW \Psi_j\dfrac{\left(\eta_{ij}+\Theta_j\right)^2-\eta_{ij}^2-\Theta_j^2}{2t} \\
    \text{ s.t. } 
    &C_1:\eta_{ij} \in \lbrace 0,1 \rbrace, \text{ } \forall i, \text{ }\forall j; \hspace{4.5cm}C_2:\sum_{j=1}^{\ngrps}\eta_{ij} \leq 1,\forall i; \nonumber \\
    &C_{3}:\sum_{i=1}^{\totUsers}\eta_{ij} \leq \delta_j \totUsers, \forall j,  
    \hspace{4.9cm}C_{4}:\delta_{j} \in \lbrace 0, 1\rbrace, \forall j; \nonumber \\
    &C_{5}:\sum_{l\neq i}|\mathbf{h}_{i}^H\mathbf{w}_l|^2+\sigma^2 \leq \mathcal{J}_{ij}(\mathbf{W},\alpha_{ij}), \text{ }\forall i, \text{ }\forall j, 
    \hspace{1.2cm}C_{6}:\sum_{j=1}^{\ntxants}\norm{\mathbf{w}_j}_2^2 \leq \totpow, \nonumber \\
    &C_7: \left(\eta_{ij}+\Theta_j\right)^2-2\log \alpha_{ij} \leq \eta_{ij}^2+\Theta_j^2  ,\text{ }\forall i, \text{ }\forall j, \hspace{0.8cm} C_8: \Theta_j \geq \delta_j \epsilon_j, \text{ } \forall j, \nonumber \\ &\color{blue}C_{9}:\text{ }\color{blue} \frac{\zeta_j}{\BW} + \delta^2_j +\Theta^2_j  \leq \left(\delta_j +\Theta_j\right)^2,\text{ }\forall j, \nonumber \\
    &\color{blue}C_{10}: P_0 + \sum_{j=1}^{\ngrps} \left(\dfrac{1}{\rho} \norm{\mathbf{w}_j}^2 + \Pi \text{ } \left({p}_{1}\left(\zeta_j\right)-{p}_{2}\left(\zeta_j\right)\right)\right)\leq t, \nonumber \\
    &C_{11}:\sum_{j=1}^{\ngrps}\delta_{j}=\ntxants, \text{ } \forall j, \nonumber 
    \end{align} 
       where $\mathcal{J}_{ij}(\mathbf{W},\alpha_{ij})\triangleq\dfrac{\sum_{l=1}^{\ngrps}|\mathbf{h}_{i}^H\mathbf{w}_l|^2+\sigma^2}{\alpha_{ij}}$, $\boldsymbol{\delta}=\left[{\delta_1},\hdots,{\delta_{\ngrps}} \right]^T$. In constraint $C_3$ in $\mathcal{P}_{3}^{\text{MEE}}$, the binary slack variable $\delta_j$ is used for controlling the scheduling of group $j$. In other words, $\delta_j=0$ indicates that group $j$ is not scheduled else scheduled. However, for scheduled group $j$ (i.e., $\delta_j=1$) constraint $C_3$ becomes superfluous as it is always satisfied. With the help $C_3$ and $C_4$, constraint $C_{10}$ ensures that number of scheduled groups is exactly $M$. \textcolor{blue}{The constraint $C_9$ in $\mathcal{P}_3^{\text{MEE}}$ is the DC reformulation of $\zeta_j \leq \BW \delta_j \Theta_j,$ $\forall j$.}

    \subsection{Continuous DC using relaxation and penalization: MEE}
    
    Ignoring the combinatorial constraints $C_1$ and $C_4$, the constraint set of $\mathcal{P}_{3}^{\text{MEE}}$ can be seen as a DC problem. So, the stationary points of such DC problems can be efficiently obtained by convex-concave procedure (CCP). With the aim of adopting the CCP framework, the binary constraints $C_1$ and $C_4$ in $\mathcal{P}_{4}^{\text{MEE}}$ are relaxed to box constraint between 0 and 1 i.e., $[0,1]$. The CCP framework can be readily applied to this relaxed continuous problem; however, the obtained stationary points might yield non-binary $\delta_j$s and $\eta_{ij}$s. Although, a quantization procedure  can be used to obtain binary $\delta_j$s and $\eta_{ij}$s, the resulting solutions  may not be even feasible to $\mathcal{P}_{1}^{\text{MEE}}$. Therefore, obtaining binary $\delta_j$s and $\eta_{ij}$ in the relaxed problem is crucial to ensure that the obtained solution are  feasible to the original problem $\mathcal{P}_2^{\text{MEE}}$. Therefore, the relaxed variables $\delta_j$s and $\eta_{ij}$s are further penalized to encourage the relaxed problem to include binary $\delta_j$s and $\eta_{ij}$s in the final solutions. Letting $\lambda_1>0$ and $\lambda_2>0$ be the penalty parameters respectively and $\mathbb{P}\left(.\right)$ be the penalty function, the penalized continuous formulation of $\mathcal{P}_{4}^{\text{MEE}}$ is,
    
    \begin{align}\label{eq:EEE_penalty_form1}
    \mathcal{P}_{4}^{\text{MEE}}:\text{} &\max_{\mathbf{W},\boldsymbol{\Theta,\eta,\delta,\alpha,\zeta } ,t } \text{ }  \sum_{i=1}^{\totUsers}\sum_{j=1}^{\ngrps}f\left(\eta_{ij},\Theta_j,t\right)+\sum_{j=1}^{\ngrps}\sum_{i=1}^{\totUsers}\lambda_1 \mathbb{P}\left(\eta_{ij}\right)+\lambda_2 \sum_{j=1}^{\ngrps}\mathbb{P}\left(\delta_j\right) \\
    \text{ s.t. } 
     &C_1:\text{} 0 \leq \eta_{ij} \leq 1 , \text{ } \forall i, \text{ }\forall j, \hspace{0.3cm} C_4:\text{ }0 \leq \delta_{j} \leq  1, \forall j, \hspace{0.3cm}
    C_2, C_3,  C_5 \text{ to } C_{11} \text{ in } \eqref{eq:EEE_MIDC} \nonumber
    \end{align} 
    
It is easy to see that any choice of convex function $\mathbb{P}\left(\eta_{ij} \right)$ that promotes the binary solutions suffice to transform $\mathcal{P}_{4}^{\text{MEE}}$ as a DC problem of our interest. The entropy based penalty function proposed in \cite{AshJSP_MISO_2019} i.e., $\mathbb{P}{\left(\eta_{ij}\right)} \triangleq \eta_{i}\log\eta_{ij}+\left(1-\eta_{ij}\right)\log\left(1-\eta_{ij}\right)$ is considered for this work. With this choice of $\mathbb{P}{\left(\eta_{ij}\right)} $, the problem $\mathcal{P}_{4}^{\text{MEE}}$ becomes a DC problem. In order to apply the CCP framework to the problem $\mathcal{P}_{4}^{\text{MEE}}$, a feasible initial point (FIP) needs to supplied. However, the constraint $C_5$ in $\mathcal{P}_{4}^{\text{MEE}}$ limits the choices of FIPs. For ease of finding the FIPs, the constraint $C_{10}$ is brought into the objective with another penalty parameter $\Omega_1 >0$ as,

 \begin{align}\label{eq:EEE_Fin_DC}
\mathcal{P}_{5}^{\text{MEE}}:\text{} &\max_{\mathbf{W},\boldsymbol{\Theta,\eta,\delta,\alpha,\zeta} ,t } \text{ }  \sum_{i=1}^{\totUsers}\sum_{j=1}^{\ngrps}f\left(\eta_{ij},\Theta_j,t\right)+\lambda_2 \sum_{j=1}^{\ngrps}\mathbb{P}\left(\delta_j\right) \nonumber \\
&+\sum_{j=1}^{\ngrps}\sum_{i=1}^{\totUsers}\lambda_1 \mathbb{P}\left(\eta_{ij}\right)-\Omega_1 \norm{\sum_{j=1}^{\ngrps}\delta_{j}-\ntxants}^2 \\
\text{ s.t. } 
&C_1 \text{ to } C_{10}\text{ in } \eqref{eq:EEE_penalty_form1} \nonumber
    \end{align}

\subsection{A CCP based Joint Design Algorithm: MEE}\label{sec:CCP_EEE}
In this section, a CCP based algorithm is proposed for joint user grouping, scheduling and precoding for MEE (JGSP-MEE) problem given in problem \eqref{eq:EEE_Fin_DC}. \textcolor{blue}{CCP  proposed in \cite{Yuille_CCCP_2001} is a special case of successive convex approximation framework \cite{SCA_org}} designed for DC programming problem. So, CCP is an iterative framework where in each iteration convexification and optimization steps are applied to the DC problem  until the convergence. The convexification and optimization steps of $\mathcal{P}_{5}^{\text{MEE}}$ of JGSP-MEE at the iteration $k$ is given as,
\begin{itemize}
\item Convexification: Let $\left({\mathbf{W}}, \boldsymbol{\eta,\text{ }\delta}, \boldsymbol{\Theta},\boldsymbol{\alpha},,t\right)^{k-1}$ be the estimates of $\left({\mathbf{W}}, \boldsymbol{\eta,\text{ }\delta}, \boldsymbol{\Theta},\boldsymbol{\alpha},t\right)$ in iteration $k-1$  respectively. In iteration $k$, the functions $\mathbb{P}\left(\delta_j\right), \mathbb{P}\left(\eta_{ij}\right), {{p}}_{2}\left(\zeta_j\right)$ and $f\left(\eta_{ij},\Theta_j,t\right)$ are replaced their first Taylor approximations $\tilde{\mathbb{P}}^{k}\left(\eta_{ij}\right),\tilde{\mathbb{P}}^{k}\left(\delta_j\right), \tilde{{p}}_{2}\left(\zeta_j\right)$, and  $f^k\left(\eta_{ij},\Theta_j,t\right)$ respectively which are given in Appendix I. Similarly, the concave parts in of $C_5$, $C_7$ and $C_9$ in $\mathcal{P}_{5}^{\text{MEE}}$ are replayed their first Taylor approximations $\tilde{\mathcal{G}}_{ij}^{k}(\eta_{ij},\Theta_j), \tilde{\mathcal{K}}_{ij}^{k}(\delta_{j},\Theta_j), \tilde{\mathcal{J}}^k_{ij}(\mathbf{W},\alpha_{ij})$ respectively given in Appendix I.
\item Optimization: Updated $\left({\mathbf{W}}, \boldsymbol{\alpha}, \boldsymbol{\Theta}, \boldsymbol{\eta},\boldsymbol{\delta}, t\right)^{k+1}$ is obtained by solving the following convex problem, 
\begin{align}\label{eq:EE_CCP_joint_convx}
  \mathcal{P}_{6}^{\text{MEE}}&:\text{}\max_{{\mathbf{W},\boldsymbol{\Theta,\zeta, \eta,\delta,\alpha} ,t}} \text{ } \sum_{i=1}^{\totUsers}\sum_{j=1}^{\ngrps} \left(f^k\left(\eta_{ij},\Theta_j,t\right)+\lambda_1 \tilde{\mathbb{P}}^k\left(\eta_{ij}\right)\right) \nonumber \\
  &\hspace{1.0cm}-\Omega_1 \norm{\sum_{j=1}^{\ngrps}\delta_{j}-\ntxants}^2+\sum_{j=1}^{\ngrps} \lambda_2 \tilde{\mathbb{P}}^k\left(\delta_j\right)  \\ 
    \text{ s.t. } 
    &C_1 \text{ to }C_{4} \text{ and } C_{6}, C_8 \text{ in } \eqref{eq:EEE_Fin_DC}, \nonumber \\
    &C_{5}:\text{ }\sum_{l\neq i}|\mathbf{h}_{j}^H\mathbf{w}_l|^2+\sigma^2 \leq  \tilde{\mathcal{J}}_{ij}^{k}({\mathbf{W}},\alpha_{ij}), \text{ }\forall i, \text{ }\forall j, \nonumber \\
    &C_{7}:\text{ }\left(\eta_{ij}+\Theta_j\right)^2 \leq 2\log\alpha_{ij}+\tilde{\mathcal{G}}_{ij}^{k}(\eta_{ij},\Theta_{j}),\text{ }\forall i,\text{ }\forall j. \nonumber \\
    &\color{blue}C_{9}:\text{ }\color{blue} \frac{\zeta_j}{\BW} + \delta^2_j +\Theta^2_j  \leq \tilde{\mathcal{K}}_{ij}^{k}(\delta_{j},\Theta_{j}),\text{ }\forall j, \nonumber \\
     &\color{blue}C_{10}: \text{ }P_0 + \sum_{j=1}^{\ngrps} \left(\dfrac{1}{\rho} \norm{\mathbf{w}_j}^2 + \Pi \text{ } \left({p}_{1}\left(\zeta_j\right)-\tilde{{p}}_{2}\left(\zeta_j\right)\right)\right)\leq t,  \nonumber 
\end{align} 
\end{itemize}

The proposed CCP based JGSP-MEE algorithm iteratively solves the problem in $\mathcal{P}_{6}^{\text{MEE}}$. However, to guarantee its convergence to a stationary point JGSP-MEE needs to be initialized with a FIP (kindly refer \cite{Sriperumbudur_CCP_2009}). In this case, $\boldsymbol{\delta}=\boldsymbol{\eta}=\boldsymbol{0}$ results a trivial FIP. Although the trivial solution is a valid FIP to the problem $\mathcal{P}_{5}^{\text{MEE}}$, it is observed through simulations that it usually converges to a poorly performing stationary point with the poor objective function value. This behavior might be due to the fact the trivial FIP has the lowest objective (i.e., zero), therefore, the JGSP-MEE initialized with the trivial FIP may converge to a stationary point around this lowest objective value. Since, FIP is crucial for JGSP-MEE's performance, in the sequel, a simple procedure is proposed to obtain a FIP that promises the convergence to stationary points which yield better performance.

\subsection{Feasible Initial Point: {MEE}}\label{sec:FIP_EE}
Since, the quality of the solution depends on the FIP, the harder task of finding a better FIP is considered through the following procedure.
\begin{itemize}
\item Step 1: Initialize $\mathbf{W}^0$ with complex random values subject to $\norm{\mathbf{W}^0}_2^2 \leq \totpow$ and calculate initial SINRs $\boldsymbol{\gamma}^0$.
\item Step 2:  Solve the following optimization: \vspace{-0.3cm}
    \begin{align}\label{eq:FIP_WSR} 
    \cal{P}_{\text{FES}}: &\text{ }\lbrace {\boldsymbol{\delta}_0,\boldsymbol{\eta}_0}\rbrace:\text{ } \max \text{ }\sum_{j=1}^{\ngrps}{\delta_j}+\sum_{j=1}^{\ngrps}\sum_{i=1}^{\totUsers}{\eta_{ij}} \\
   \text{s.t. } 
    &C_1:\text{ } 0 \leq \eta_{ij} \leq 1, \text{ } \forall i, \text{ }\forall j,
    &&C_2:\text{} \sum_{j=1}^{\ngrps}\eta_{ij} \leq 1,\forall i, \nonumber \\
    &C_{3}:\text{ }\sum_{i=1}^{\totUsers}\eta_{ij} \leq  \delta_j \totUsers, \forall j, 
    &&C_{4}:\text{ } 0 \leq \delta_{j} \leq 1, \forall j, \nonumber \\
    &C_{5}:\text{ }\log\left(1+\gamma_{ij}^0\right) \geq \eta_{ij}{\epsilon_j},\text{ }\forall i,\text{ }\forall j, \nonumber 
    \end{align}
\item Step 3: The parameters $\boldsymbol{\Theta}^0,\boldsymbol{\zeta}^0, \boldsymbol{\alpha}^0, t^0$ can easily be derived from $\mathbf{W}^0, \boldsymbol{\delta}^0$ and $\boldsymbol{\eta}^0$.
\end{itemize}
\emph{Remarks:}
   \begin{itemize}
   \item \textcolor{blue}{The problem $\cal{P}_{\text{FES}}$ is a linear programming problem and always feasible since trivial solution $\boldsymbol{\delta}^0=\boldsymbol{\eta}^0=\boldsymbol{0}$ is also a feasible solution. However, the optimization problem $\cal{P}_{\text{FES}}$ usually results a better solution than  trivial one. Therefore, initial parameters $\mathbf{W}^0, \boldsymbol{\delta}^0,\boldsymbol{\Theta}^0,\boldsymbol{\eta}^0,\boldsymbol{\alpha}^0,t^0$ are always feasible. Different $\mathbf{W}^0$ in step 1 may lead to different FIPs.}
   \item The optimization problem in Step 2 is a linear programming problem which can be solved efficiently to large dimensions with many of the existing tools like CVX.
   \item The FIP obtained by this procedure may not be feasible for the original MEE problem $ \cal{P}_1^{\text{MEE}}$ unless $\mathbf{W}^0,\boldsymbol{\eta}^0,\boldsymbol{\delta}^0$ becomes feasible to $\cal{P}_2^{\text{MEE}}$.
   \item Although the FIP obtained by this method is not feasible for $\mathcal{P}_{1}^{\text{MEE}}$, the final solution obtained by JGSP-MEE with this FIP becomes a feasible for $\mathcal{P}_{1}^{\text{MEE}}$ since the final solution satisfies the group scheduling constraint $C_2$ in $\mathcal{P}_{1}^{\text{MEE}}$.
   \end{itemize}
   Letting $\cal{P}_6^{\text{MEE}}\left(k\right)$ be the objective value of the problem $\cal{P}_6^{\text{MEE}}$ at  iteration $k$, the pseudo code of JGSP-MEE for the joint design problem is given in algorithm~\ref{alg:JGSP_EEE}.
\begin{algorithm}
\setlength{\textfloatsep}{0pt}
 \caption{JGSP-MEE}
 \label{alg:JGSP_EEE}
 \begin{algorithmic}[]
 \State{\textbf{Input}: $\mathbf{H},\left[\epsilon_1,\hdots,\epsilon_{\totUsers}\right],\totpow,\Delta, \mathbf{W}^0, \boldsymbol{\delta}^0,\boldsymbol{\Theta}^0,\boldsymbol{\zeta}^0, \boldsymbol{\eta}^0,\boldsymbol{\alpha}^0,t^0$, $\lambda_1=0, k=1$}; \\
 \textbf{Output}: $\mathbf{W},\boldsymbol{\eta}$
 \While{$|\cal{P}_6^{\text{MEE}}\left(k\right)-\cal{P}_6^{\text{MEE}}\left(k-1\right)|\geq \Delta$}
    \State \textbf{Convexification:} Convexify the problem \eqref{eq:EEE_Fin_DC}
    \State \textbf{Optimization}: Update $\left({\mathbf{W}}, \boldsymbol{\eta,\delta,\alpha,\zeta,\Theta},t\right)^{k}$ by solving $\mathcal{P}_{6}^{\text{MEE}}$
    \State \textbf{Update :} $\cal{P}_6^{\text{MEE}}\left(k\right), \lambda_1,\lambda_2,\Omega_1, k$
 \EndWhile
\end{algorithmic}
\end{algorithm}  

   

   

\subsection{Complexity of JGSP-MEE} Since JGSP-MEE is a CCP based iterative algorithm, its complexity depends on complexity of the convex sub-problem $\mathcal{P}_{6}^{\text{MEE}}$. The convex problem $\mathcal{P}_{6}^{\text{MEE}}$ has $\left(\ntxants\ngrps+2\totUsers \ngrps +3\ngrps+1\right)$ decision variables and $\left(2\totUsers\ngrps + 2+\ngrps\right)$ convex constraints and $\left(2\totUsers\ngrps+4\ngrps+\totUsers\right)$ linear constraints. Hence, the complexity of  $\mathcal{P}_{6}^{\text{MEE}}$ is $\mathcal{O}\left(\left(\ntxants\ngrps+2\totUsers \ngrps +2\ngrps+1\right)^3 \left(4\totUsers\ngrps+4\ngrps+\totUsers+2\right)\right)$ \cite{Complexity_ref}. Commercial software such as CVX can solve the convex problem of type $\mathcal{P}_{6}^{\text{MEE}}$ efficiently to a large dimension. Besides the  complexity per iteration, the overall complexity also depends on the convergence speed of the algorithm. Through simulations, we observe that the JGSP-MEE converges  typically in 15-20 iterations. 
\section{Variants of Multicast Energy Efficiency}\label{sec:EE}
In this section, two special cases of the MEE problem namely the maximization of EE and the number of scheduled users are considered.
\subsection{Energy efficiency}
In this section, we focus on developing a CCP based low-complexity algorithm for the joint design of user grouping, scheduling, and precoding for maximization of weighted EE (defined in \eqref{eq:EE_def}) subject to grouping, scheduling, precoding, power, and QoS constraints. This problem is simply referred to as the EE problem.
\subsubsection{Problem formulation: EE}
With the defined slack variables in Section~\ref{sec:MEE}, the EE problem is mathematically formulated as,
\begin{align}\label{eq:EE_SR_MINLP1}
    \mathcal{P}_{1}^{\text{EE}}:\text{} &\max_{\mathbf{W},\boldsymbol{\Theta,\eta,\delta,\alpha} } \text{ }   \dfrac{ \sum_{j=1}^{\ntxants} \BW \Psi_j \Theta_j}{t}   \\
    \text{ s.t. } 
   & C_1 \text{ to }  C_{8} \text{ and } C_{11} \text{ in } \eqref{eq:EEE_MIDC}, \nonumber \\ 
   &\color{blue}C_{9}: \text{ }P_0 + \sum_{j=1}^{\ngrps} \left(\dfrac{1}{\rho} \norm{\mathbf{w}_j}^2 + \Pi \text{ } {p}\left(\Theta_j\right)\right)\leq t,  
   \hspace{0.5cm} C_{10}:  \Theta_j \leq \delta_j \Theta^*, \text{} \forall j,  \nonumber
    \end{align} 
 where $\Theta^*\geq \max_{j=1}^{\ngrps} \Theta_j$ is a constant. The constant $\Theta^*$ in constraint $C_{12}$ in $\mathcal{P}_{1}^{\text{EE}}$ is used for forcing $\Theta_j$ to zero  when the group is not scheduled i.e., $\delta_j=0$. For the scheduled group i.e., $\delta_j=1$ constraint $C_{12}$ becomes superficial as $\Theta_j \leq \Theta*$ is always true. Without the constraint $C_{12}$ the problem $\mathcal{P}_{1}^{\text{EE}}$ becomes unbounded as the $\Theta_j$ can be infinity for the unscheduled group $j$ thus yielding the highest EE which is infinity. The constraint $C_{12}$ helps in containing $\Theta_j$ to zero for the unscheduled group $j$. Therefore the problem $\mathcal{P}_{1}^{\text{EE}}$ becomes bounded due to $C_{12}$. \textcolor{blue}{Notice the difference between the constraint $C_9$ in $\mathcal{P}_1^{\text{EE}}$ and $C_{10}$ in $\mathcal{P}_3^{\text{MEE}}$. Due to $C_{10}$ in $\mathcal{P}_1^{\text{EE}}$ for an unscheduled group $j$ the minimum rate of the group i.e., $\Theta_j$ is zero. Therefore, the power consumption can be modelled simply using $\Theta_j$ unlike $\zeta_j$ in MEE case.}


\paragraph*{Nature of EE in the context of grouping and scheduling}
EE problem is not biased to favor the solutions with more number of users since it only considers the minimum rate of the group ignoring its size. Typically, adding more users to groups either leads to increased inter-group interference and/or lower minimum rate of the group due to lower channel gains. Hence, to obtain the same rate as with few users extra power needs to be used. Since the linear increase in rate is achieved at the cost of exponential increase power, newly added users result in lower EE.

\subsubsection{DC formulation and CCP based algorithm: EE}
 The problem $\mathcal{P}_{1}^{\text{EE}}$ is combinatorial and non-convex similar to the problem $\mathcal{P}_{3}^{\text{MEE}}$. With the help of a slack variable $\Gamma$, and applying reformulations and relaxations proposed in Section~\ref{sec:MEE}, the problem $\mathcal{P}_{1}^{\text{EE}}$ is reformulated into a DC problem as,
  \begin{align}\label{eq:EE_MIDC}
    \mathcal{P}_{2}^{\text{EE}}:\text{} &\max_{\mathbf{W},\boldsymbol{\Theta,\eta,\delta,\alpha},\Gamma,t}\text{ }
      \dfrac{  \Gamma^2}{t} -{\Omega}_2 \norm{\sum_{j=1}^{\ngrps}\delta_{j}-\ntxants}^2  +\sum_{j=1}^{\ngrps}\sum_{i=1}^{\totUsers}{\lambda}_3 \mathbb{P}\left(\eta_{ij}\right)+{\lambda}_4 \sum_{j=1}^{\ngrps}\mathbb{P}\left(\delta_j\right) \\
    \text{ s.t. } 
   & C_1 \text{ to }  C_{10}\text{ in } \eqref{eq:EE_SR_MINLP1}, \hspace{1cm} C_{11}:  \sum_{j=1}^{\ngrps}\BW \Psi_j \Theta_j \geq \Gamma^2. \nonumber
    \end{align} 
where ${\lambda}_3>0, {\lambda}_4>0$ and ${\Omega}_2>0$ are the penalty parameters. 

Notice that the DC problem $\mathcal{P}_{2}^{\text{EE}}$ resembles the DC problem $\mathcal{P}_{5}^{\text{MEE}}$, hence, the CCP framework proposed in Section~\ref{sec:CCP_EEE} can be simply be adapted. The proposed CCP framework based algorithm for the EE problem is simply referred to as JGSP-EE. Since JGSP-EE is a CCP based iterative algorithm at iteration $k$ it executes the following convex problem:
\begin{align}\label{eq:EE_SR_CCP_joint_convx}
  &\mathcal{P}_{3}^{\text{EE}}:\text{}\max_{{\mathbf{W},\boldsymbol{\Theta,\eta,\delta,\alpha},\Gamma ,t}} \text{ }   \dfrac{  2\Gamma^{k-1}\Gamma}{t} -{\Omega}_2 \norm{\sum_{j=1}^{\ngrps}\delta_{j}-\ntxants}^2 +{\lambda}_3 \sum_{j=1}^{\ngrps} \sum_{i=1}^{\totUsers} \tilde{\mathbb{P}}^k\left(\eta_{ij}\right)+\sum_{j=1}^{\ngrps} {\lambda}_4 \tilde{\mathbb{P}}^k\left(\delta_j\right)  \\ 
    &\text{ s.t. } 
    C_1 \text{ to } C_{8} \text{ in } \eqref{eq:EEE_Fin_DC},
     \hspace{1.7cm}\color{blue}C_{9}: \text{ }P_0 + \sum_{j=1}^{\ngrps} \left(\dfrac{1}{\rho} \norm{\mathbf{w}_j}^2 + \Pi \text{ } \left( {p}_1\left(\Theta_j\right)- \tilde{{p}}_2\left(\Theta_j\right)\right)\right)\leq t,  \nonumber \\
    &\hspace{0.7cm} C_{10}:  \Theta_j \leq \delta_j \Theta^*, \text{} \forall j,  \hspace{1.2cm}C_{11}:  \sum_{j=1}^{\ngrps}\BW \Psi_j \Theta_j \geq \Gamma^2. \nonumber
\end{align} 
Letting $\mathcal{P}_{2}^{\text{EE}}\left(k\right)$ be the objective value of the problem $\mathcal{P}_{2}^{\text{EE}}$ at  iteration $k$, the pseudo code of JGSP-EE for the joint design problem is given in algorithm~\ref{alg:JGSP_EE_SR}.
\begin{algorithm}
 \caption{JGSP-EE-SR}
 \label{alg:JGSP_EE_SR}
 \begin{algorithmic}[]
 \State{\textbf{Input}: $\mathbf{H},\left[\epsilon_1,\hdots,\epsilon_{\totUsers}\right],\totpow,\Delta, \mathbf{W}^0, \boldsymbol{\delta}^0,\boldsymbol{\Theta}^0,\boldsymbol{\eta}^0,\boldsymbol{\alpha}^0,t^0$, $\lambda_3,\lambda_4,\Omega_2, k=1$}; \\
 \textbf{Output}: $\mathbf{W},\boldsymbol{\eta}$
 \While{$|\mathcal{P}_{2}^{\text{EE}}\left(k\right)\mathcal{P}_{2}^{\text{EE}}\left(k-1\right)\left(k-1\right)|\geq \Delta$}
    \State \textbf{Convexification:} Convexify the problem \eqref{eq:EEE_Fin_DC}
    \State \textbf{Optimization}: Update $\left({\mathbf{W}}, \boldsymbol{\eta,\delta,\alpha,\Theta},\Gamma, t\right)^{k}$ by solving $\mathcal{P}_{2}^{\text{EE}}\left(k\right)$
    \State \textbf{Update :} $\mathcal{P}_{2}^{\text{EE}}\left(k\right), {\lambda}_3,{\lambda}_4,{\Omega}_2, k$
 \EndWhile
\end{algorithmic}
\end{algorithm}

 \subsection{Maximization of scheduled users}\label{sec:SUM}
 In this section, the problem of maximizing the scheduled users (SUM) is considered subject to grouping, scheduling, precoding, total power, and QoS constraints. This problem is simply referred to as SUM problem in this paper and is formulated as, 
 
  \begin{align}\label{eq:SUM_Bin_form1}
    \mathcal{P}_{1}^{\text{SUM}}:\text{} &\max_{\mathbf{W},\boldsymbol{\Theta,\eta,\delta} } \text{ }  \sum_{i=1}^{\totUsers}\sum_{j=1}^{\ngrps} \eta_{ij} \\
    \text{ s.t. } 
    C_1:&\text{ } \eta_{ij} \in \lbrace 0,1 \rbrace, \text{ } \forall i, \text{ }\forall j,
    && C_2:\text{ } \sum_{j=1}^{\ngrps}\eta_{ij} \leq 1,\forall i, 
    &&&C_{3}:\text{ }\sum_{i=1}^{\totUsers}\eta_{ij} \leq \delta_j \totUsers, \forall j, \nonumber \\
    C_{4}:&\text{ }\delta_{j} \in \lbrace 0, 1\rbrace, \forall j, 
    &&C_{5}:\text{ }\sum_{j=1}^{\ngrps}\delta_{j}=\ntxants, \text{ } \forall j, 
    &&&C_6:\text{ }\sum_{i=1}^{\ntxants}\norm{\mathbf{w}_i}_2^2 \leq \totpow,  \nonumber \\
    C_{7}:& \text{ }1+\gamma_{ij} \geq 1+\eta_{ij}\epsilon_j, \text{ }\forall i, \text{ }\forall j. \nonumber
    \end{align} 
    
Notice that except for constraint $C_6$ all the constraints and the objective in $\mathcal{P}_{1}^{\text{SUM}}$ are linear and convex. Further, similar to the constraint $C_4$ in $\mathcal{P}_2^{\text{EE}}$, the constraint $C_6$ can be easily equivalently transformed as a DC. Therefore, with the help of the relaxations and penalization approach provided in Section~\ref{sec:MEE} and ~\ref{sec:EE}, the problem can be transformed as a DC programming problem. Hence, the CCP framework can be adapted to solve the resulting DC problem. The transformed DC problem and the convexified problem to be solved in the CCP framework for the SUM problem are given appendix~\ref{sec:appnedix}. The CCP framework based algorithm proposed for the SUM problem is simply referred to as JGSP-SUM.

\section{Simulation results}\label{sec:simulation}
\subsection{Simulation setup and parameter initialization}\label{sec:SimSet}
\subsubsection*{Simulation setup}
In this section, the performance of the proposed algorithms JGSP-MEE, JGSP-EE and JGSP-SUM is evaluated. The system parameters discussed in this paragraph are common for all the figures. Bandwidth for all the groups is assumed to be 1 Hz i.e., $\BW=1$ Hz. The coefficients of the channel matrix, i.e., $h_{ij}$ are drawn from the complex normal distribution with zero mean and unit variance and noise variances at the receivers are considered to be unity i.e., $\sigma^2=1$. All the simulation results are averaged over 100 different channel realizations (CRs). Weights are assumed to be unity i.e., $\lbrace\Psi_j=1\rbrace_{j=1}^{\ngrps}$. Following are the acronyms/definitions commonly used for all simulation results: 1) \emph{Number of scheduled users}: the sum of all the scheduled users in all scheduled groups. 2) \emph{Orthogonal user}: An user with zero channel correlations  with all the users in all the other groups. 3) \emph{Non-orthogonal user}: A user with at least one non-zero channel correlation with any user in other groups. 4) \emph{Consumed power} $ \triangleq P_0 + \sum_{j=1}^{\ngrps} \left(\dfrac{1}{\rho} \norm{\mathbf{w}_i}^2 + \Pi {p}\left(r_j\right)\right)$. 5) \emph{Throughput} $\triangleq \sum_{j=1}^{\ntxants}  \BW \Theta_j $.

\subsubsection*{Parameter initialization} power amplifier efficiency i.e., $\rho$ is assumed to 0.2 and fixed static power i.e., $P_0$ is assumed to be 16 Watts, $\BW=1$MHz, $\Pi=2.4 {\text{Watts}}/{\left({\text{bits}}/{\text{sec}}\right)^2}$ \cite{EE_Oskari}, and $p\left(x\right)=x^2$ \cite{Yang_EE}. The penalty parameters responsible for binary nature of $\boldsymbol{\eta,\text{ }\delta}$ are initialized as follows $\lambda_1=\lambda_2=0.01$ and , $\lambda_3=\lambda_4=0.5$ and $\lambda_5=\lambda_6=0.05$ and . Further $\lbrace\lambda_i\rbrace_{i=1}^{6}$ are incremented by factor 1.2. Further, penalty parameters corresponding to group scheduling constraint are initialized to relatively larger values such as  $\Omega_1=2.5$, $\Omega_2=5$, and $\Omega_3=1$ and are incremented by 1.5 in each iteration. MEE and SUM maximization criteria naturally encourage the solutions towards to non-zero $\boldsymbol{\eta}$ and $\boldsymbol{\delta}$. Therefore, small initial values and slow update of penalty parameters corresponding to  MEE and SUM problems eventually result in a binary solution of $\boldsymbol{\eta,\text{ }\delta}$. Further, relatively large initial value and larger increments for $\Omega_1$ and $\Omega_2$ in each iteration, along with binary nature of $\boldsymbol{\delta}$, eventually ensure the group scheduling constraint i.e., $\sum_{j=1}^{\ngrps}\delta_j=\ntxants$. On the contrary, as discussed in Section~\ref{sec:EE}, the EE problem is not biased to favor the solutions with a higher number of users since it only considers the minimum rate of the group ignoring its size. Moreover, the solutions with $\eta_{ij}<1$ might be encouraged as it would facilitate larger $\Theta_j$ hence better objective in $\mathcal{P}_1^{\text{EE}}$ (from constraint $C_5$ in the problem $\mathcal{P}_1^{\text{EE}}$). Hence, to ensure the group scheduling constraint and the binary nature of $\boldsymbol{\eta,\delta}$, the penalty parameters are initialized to relatively larger values in EE than in MEE and SUM problems and incremented in large steps.

\subsection{Performance as a function of total users ($\totUsers$)}\label{sec:Perf_N}
 \begin{figure}[!htp]
\centering
    \subfloat[]{\includegraphics[height=6cm,width=8cm]{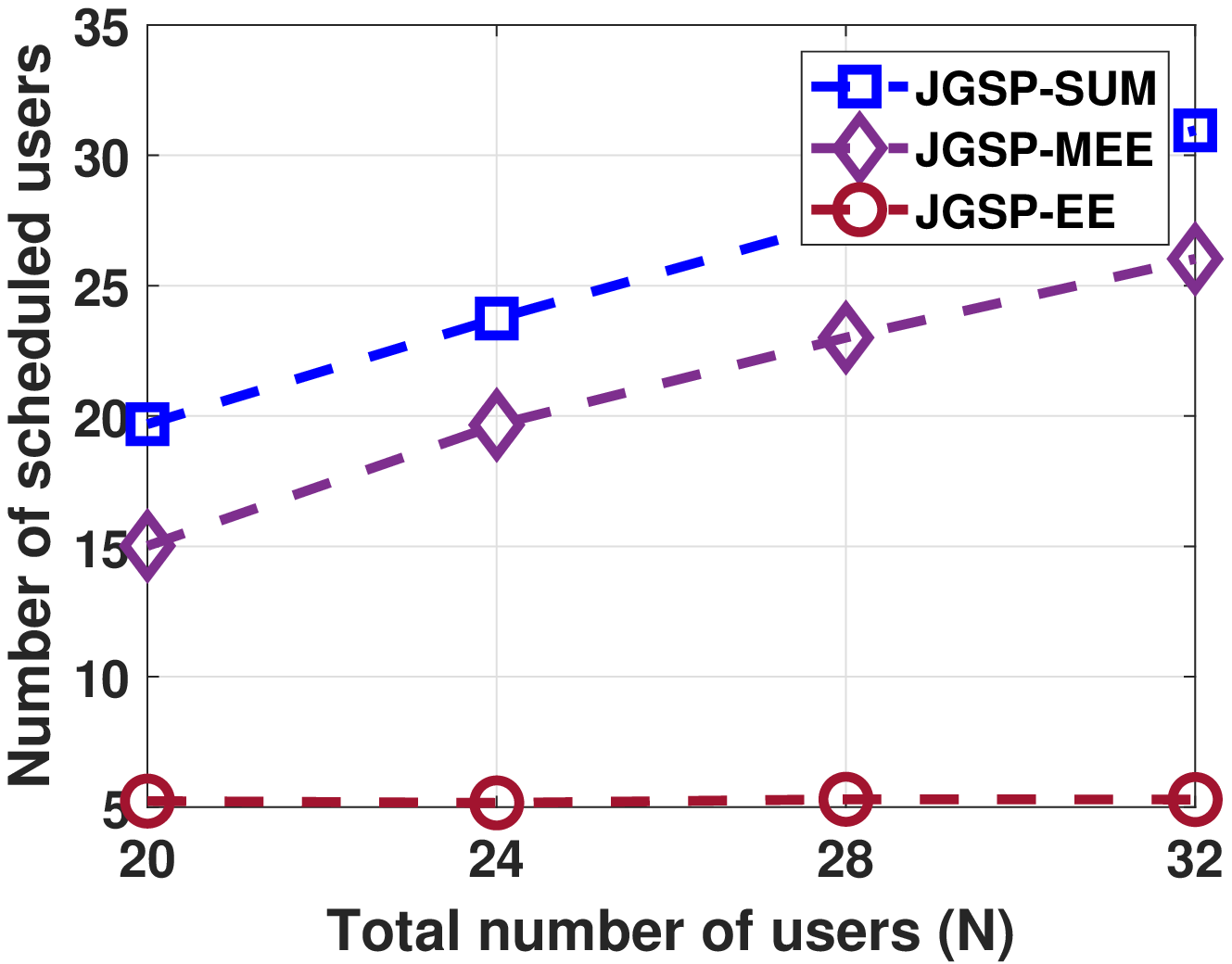} \label{fig:NSU_vs_N}}
    \subfloat[]{\includegraphics[height=6cm,width=8cm]{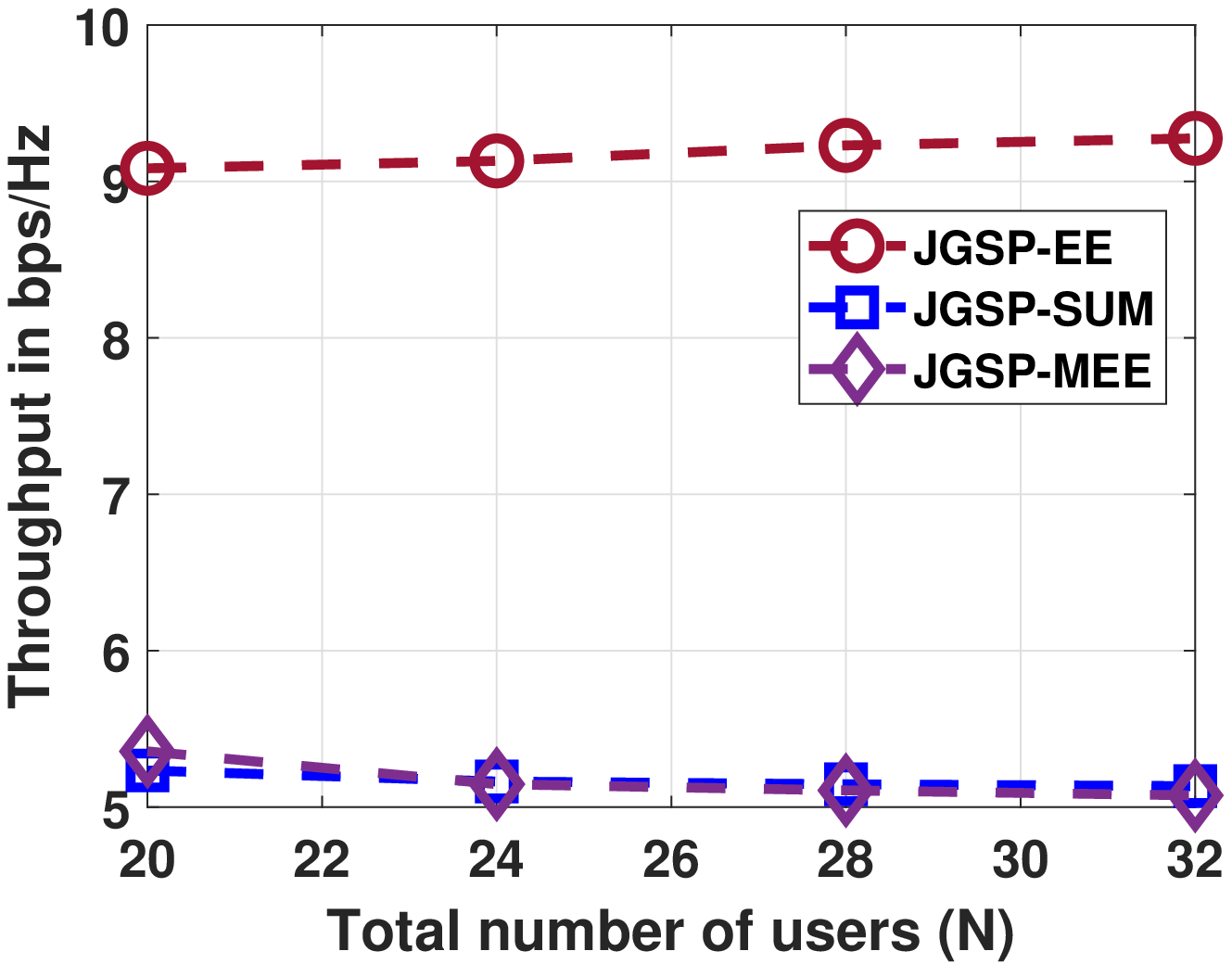}\label{fig:SE_vs_N}}\\
    \subfloat[]{\includegraphics[height=6cm,width=8cm]{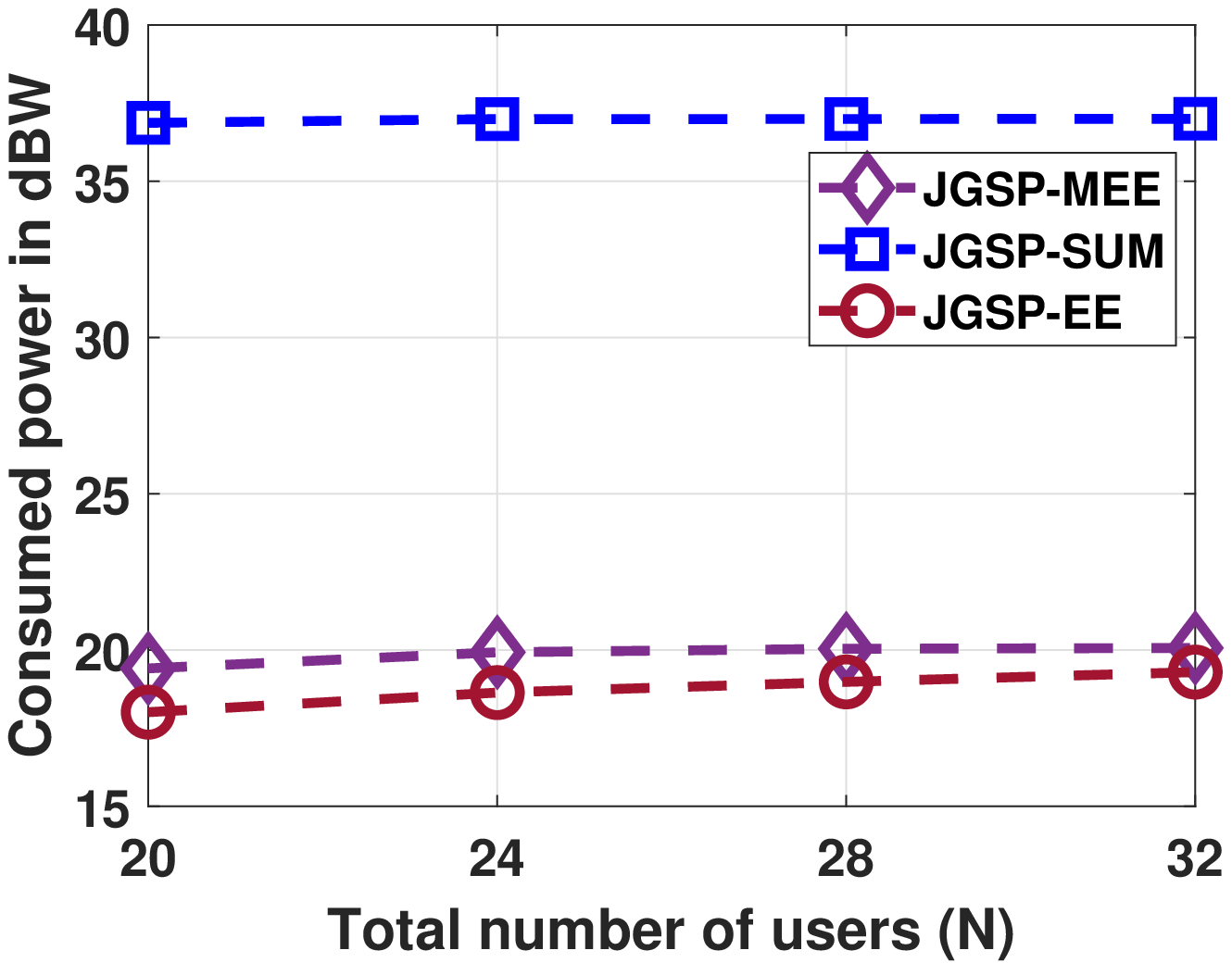}\label{fig:Pow_vs_N}}
     \subfloat[]{\includegraphics[height=6cm,width=8cm]{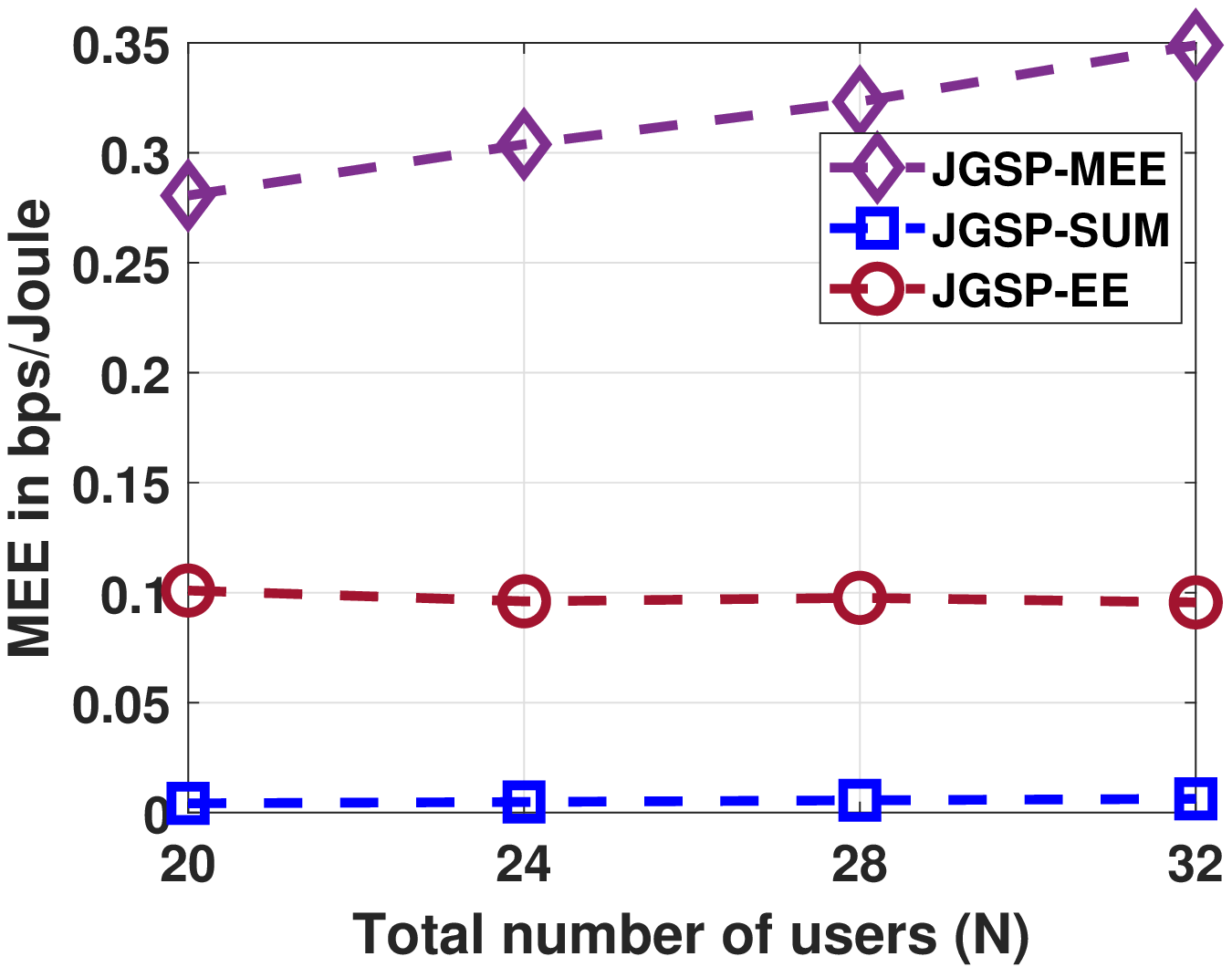}\label{fig:MEE_vs_N}}
    \caption{Comparison of proposed algorithms as a function of $\totUsers$ varying from 20 to 32 for $\ntxants=5, \ngrps=8$, $\lbrace {\epsilon}_i =1\text{bps/Hz}\rbrace_{j=1}^{\totUsers}$, and $\totpow=20\text{dBW}$ (a) number of scheduled users versus $\totUsers$ (b) throughput in bps versus $\totUsers$ (c) Consumed power in dBW versus $\totUsers$ (d) MEE in bits/Joule versus $\totUsers$}\label{fig:Perf_vs_N}
\end{figure}
In figure \ref{fig:Perf_vs_N}, the performance of the proposed algorithm i.e., JGSP-MEE, JGSP-EE, and JGSP-SUM is illustrated as a function of $N$ varying from 20 to 32 in steps of 4 for $M=5$, $\ngrps=8$, $\totpow=10$dBW and ${\epsilon}_j=1$ bps/Hz, $\forall j$. 
\subsubsection{Number of scheduled users versus $\totUsers$}
In figure~\ref{fig:NSU_vs_N}, the number of scheduled users is illustrated as function of $\totUsers$. Since JGSP-SUM directly maximizes the number of scheduled users, it schedules the maximum number of users compared to JGSP-MEE and JGSP-EE. Moreover, due to low QoS requirement and availability of resources to satisfy the QoS requirement, JGSP-SUM schedules almost all the users despite the increase in $N$. Since the number of scheduled users contribute linearly to MEE objective a similar increase in the number of scheduled users versus $N$ in JSP-MEE can be observed in figure~\ref{fig:NSU_vs_N}. However, JSP-MEE also considers the power consumed by the scheduled users, hence, JSP-MEE schedules fewer users than JSP-SUM as scheduling these excess users requires huge power which can be observed in figure~\ref{fig:Pow_vs_N}.
On the contrary, the EE objective is not accounting for the number of scheduled users, hence, JGSP-EE schedules the lowest number of users i.e. $\ntxants=5$. In other words, it is serving one user per group which is nothing but a unicast scenario. Furthermore, despite the increase in $\totUsers$, the number of users scheduled by JGSP-EE remains the same. This can be attributed to three reasons: 1) non-orthogonal users: scheduling any non-orthogonal user increases interference to users in other groups which decreases the minimum rate of the influenced groups hence decreases EE. 2) Orthogonal users with un-equal channel gains: EE swaps the existing user with the best available user in the pool as scheduling the second best user decreases the minimum rate of the group hence lower EE. 3) Orthogonal users with equal channel gains: This is an unlikely event; even if such users exist, as mentioned earlier, their scheduling is not guaranteed as the EE objective is unaffected. 

\subsubsection{Throughput versus $\totUsers$}
In figure~\ref{fig:SE_vs_N}, the throughput in bps obtained by JGSP-MEE, JGSP-EE, and JGSP-SUM is illustrated as a function of $\totUsers$.  The nature of JSP-MEE to schedule more users and consume fewer power results in lower throughput than JSP-SUM and JSP-EE. On the other hand, as the JSP-EE objective includes throughput in the objective, hence, it naturally achieves higher throughput than JSP-SUM. Moreover, as $\totUsers$ increases the probability of finding $\ntxants$ orthogonal users with good channels increases. This leads to a better throughput in JSP-EE with an increase of $N$. However, the gains in throughput for JGSP-EE diminishes as the gains in multiuser diversity diminish. On the contrary, an increase in multiuser diversity with $N$ is utilized to schedule a higher number of users by JSP-MEE and JSP-SUM which can be observed in figure~\ref{fig:NSU_vs_N}.
Moreover, the degradation in throughput in JSP-MEE and JSP-SUM is due to the combination of two factors: 1) for relatively lower $\totUsers$ i.e., 20, after scheduling the maximum number of users, resources could be used to improve minimum throughput of the groups. 2) for relatively higher $\totUsers$, as scheduling higher users improve the objectives of JSP-SUM and JSP-EEE, the available power is used to schedule more users and this is also achieved by keeping their achieved minimum rate close to the required rates of the groups. Hence, the throughput by JSP-MEE and JSP-SUM decreases slightly with an increase of $N$. 

\subsubsection{Consumed power versus $\totUsers$}
In figure~\ref{fig:SE_vs_N}, the consumed power in Watts by JSP-SUM, JSP-EE and JSP-MEE is illustrated as a function of $\totUsers$. As the JSP-SUM does not optimize power, in the process of scheduling the maximum number of users (as shown in figure~\ref{fig:NSU_vs_N}) it inefficiently utilizes the power by consuming all of the available power as depicted in figure~\ref{fig:Pow_vs_N}. On the contrary, as the EE and MEE objectives are penalized inversely for excess usage of power, both JSP-EE and JSP-MEE utilize power efficiently as shown in figure~\ref{fig:Pow_vs_N}. However, JSP-MEE slightly utilizes more power than JSP-EE as illustrated in figure~\ref{fig:Pow_vs_N} to schedule a higher number of users (as shown in figure~\ref{fig:NSU_vs_N}) as it improves over the MEE.

\subsubsection{MEE versus $\totUsers$} Recall that MEE can be interpreted as the number of received bits for one joule of transmitted energy as explained in Section~\ref{sec:EEE_def}. It can be seen in figure~\ref{fig:MEE_vs_N}, by directly optimizing MEE,  JGSP-MEE obtains the highest MEE value compared to JGSP-EE and JGSP-SUM. The linear increase in MEE with respect to $N$ can be observed in JSP-MEE and JSP-SUM as the number of scheduled users linearly with $N$ in both the methods. However, as JSP-SUM utilizes the power inefficiently, it results in poorer MEE overall compared to JSP-MEE. Unlike JSP-MEE and JSP-SUM, the improvement in MEE obtained by JSP-EE is negligible as it does not gain in scheduled users and the increase in throughput is comparatively negligible.
\subsubsection{\color{blue}Convergence of JSP-MEE versus iterations}
\begin{figure}[!htp]
    \setlength\intextsep{0pt}
    \centering
\includegraphics[height=7.75cm,width=10.0cm]{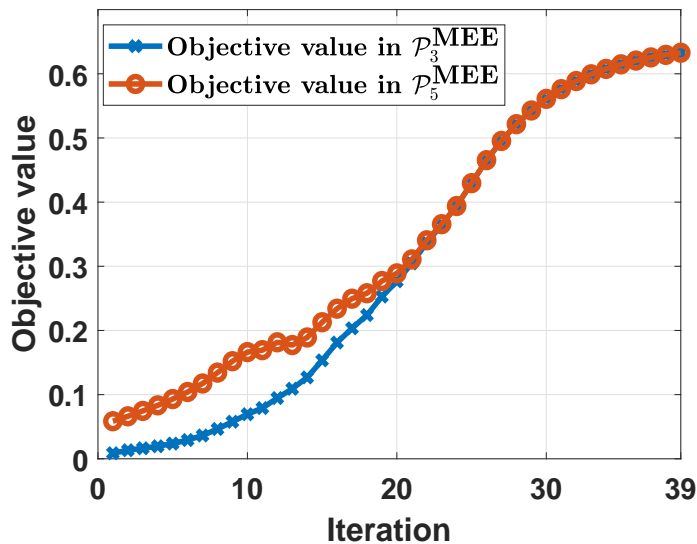}
    \caption{\color{blue}Convergence of JSP-MEE as a function of iterations for $\ntxants=5, \ngrps=8$, $\lbrace {\epsilon}_i =1\text{bps/Hz}\rbrace_{j=1}^{\totUsers}$, $\totUsers=40$ and $\totpow=30$ dBW}\label{fig:Convergence}
\end{figure}
\textcolor{blue}{
Figure~\ref{fig:Convergence} illustrates the convergence behavior of the proposed algorithm as function of iteration. The objective value in $\mathcal{P}_3^{\text{MEE}}$ is simply the MEE value i.e., $\sum_{i=1}^{\totUsers}\sum_{j=1}^{\ngrps}f\left(\eta_{ij},\Theta_j,t\right)$ and the objective value in $\mathcal{P}_5^{\text{MEE}}$ contains the MEE value plus the penalty values added to ensure binary nature of $\boldsymbol{\eta}, \boldsymbol{\delta}$ and GSC constraint i.e., $ \sum_{i=1}^{\totUsers}\sum_{j=1}^{\ngrps}f\left(\eta_{ij},\Theta_j,t\right)+\lambda_2 \sum_{j=1}^{\ngrps}\mathbb{P}\left(\delta_j\right) +\sum_{j=1}^{\ngrps}\sum_{i=1}^{\totUsers}\lambda_1 \mathbb{P}\left(\eta_{ij}\right)-\Omega_1 \norm{\sum_{j=1}^{\ngrps}\delta_{j}-\ntxants}^2$. In the initial iterations, $\boldsymbol{\eta}, \boldsymbol{\delta}$ and GSC constraints are not satisfied, hence, $\mathcal{P}_5^{\text{MEE}}$ has higher objective value than $\mathcal{P}_3^{\text{MEE}}$ which can be observed in Figure~\ref{fig:Convergence} until iteration 18. However, from iteration 19  the objective value of $\mathcal{P}_5^{\text{MEE}}$ and $\mathcal{P}_3^{\text{MEE}}$ almost same. This is because the additional penalty objective in $\mathcal{P}_5^{\text{MEE}}$ becomes zero i.e., $+\lambda_2 \sum_{j=1}^{\ngrps}\mathbb{P}\left(\delta_j\right) +\sum_{j=1}^{\ngrps}\sum_{i=1}^{\totUsers}\lambda_1 \mathbb{P}\left(\eta_{ij}\right)-\Omega_1 \norm{\sum_{j=1}^{\ngrps}\delta_{j}-\ntxants}^2=0$ as the binary nature of $\boldsymbol{\eta, \delta}$ and GSC constraints are satisfied by iteration 19. The proposed algorithm converges in 39 iterations for the system with $\totUsers=40$, $\ntxants=5$ and $\ngrps=8$. In other words, the proposed algorithm JSP-MEE exhibits the linear convergence rate which can be observed in Figure~\ref{fig:Convergence}.}

\subsubsection{\color{blue}Number of scheduled users versus versus MEE} 
\begin{figure}[!htp]
    \setlength\intextsep{0pt}
    \centering
\includegraphics[height=7.75cm,width=10.0cm]{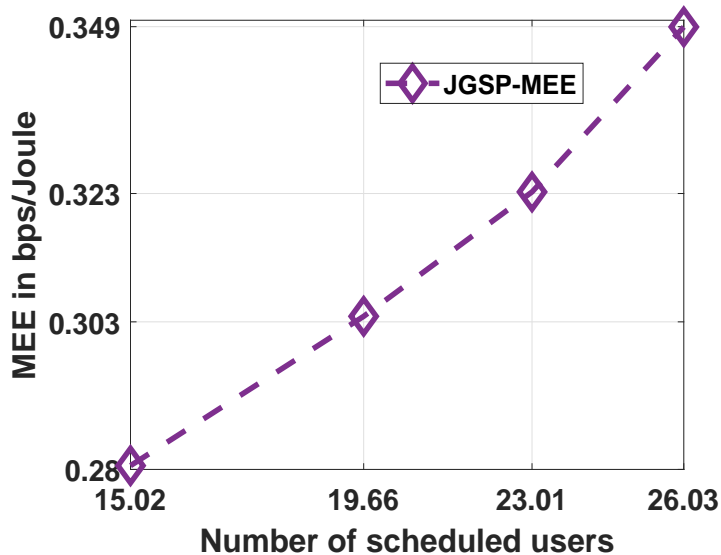}
    \caption{\color{blue}Performance of JSP-MEE as function of Number of scheduled users versus MEE for $\totUsers$ varying from 20 to 32, $\ntxants=5, \ngrps=8$, $\lbrace {\epsilon}_i =1\text{bps/Hz}\rbrace_{j=1}^{\totUsers}$, and $\totpow=30$ dBW}\label{fig:MEE_vs_NSU}
\end{figure}

\textcolor{blue}{In figure~\ref{fig:MEE_vs_NSU}, MEE obtained by JSP-MEE is plotted as function of the number of users scheduled by JSP-MEE. The linear increase in MEE of JSP-MEE with respect to the number of scheduled users is observed in figure~\ref{fig:MEE_vs_NSU}. In other words, figure~\ref{fig:MEE_vs_NSU} confirms that major contributing factor to MEE maximization is the number of scheduled users.}

\subsection{Performance as a function of total power $\totpow$}
\begin{figure}[!htp]
\centering
\subfloat[]{\includegraphics[height=6.15cm,width=8cm]{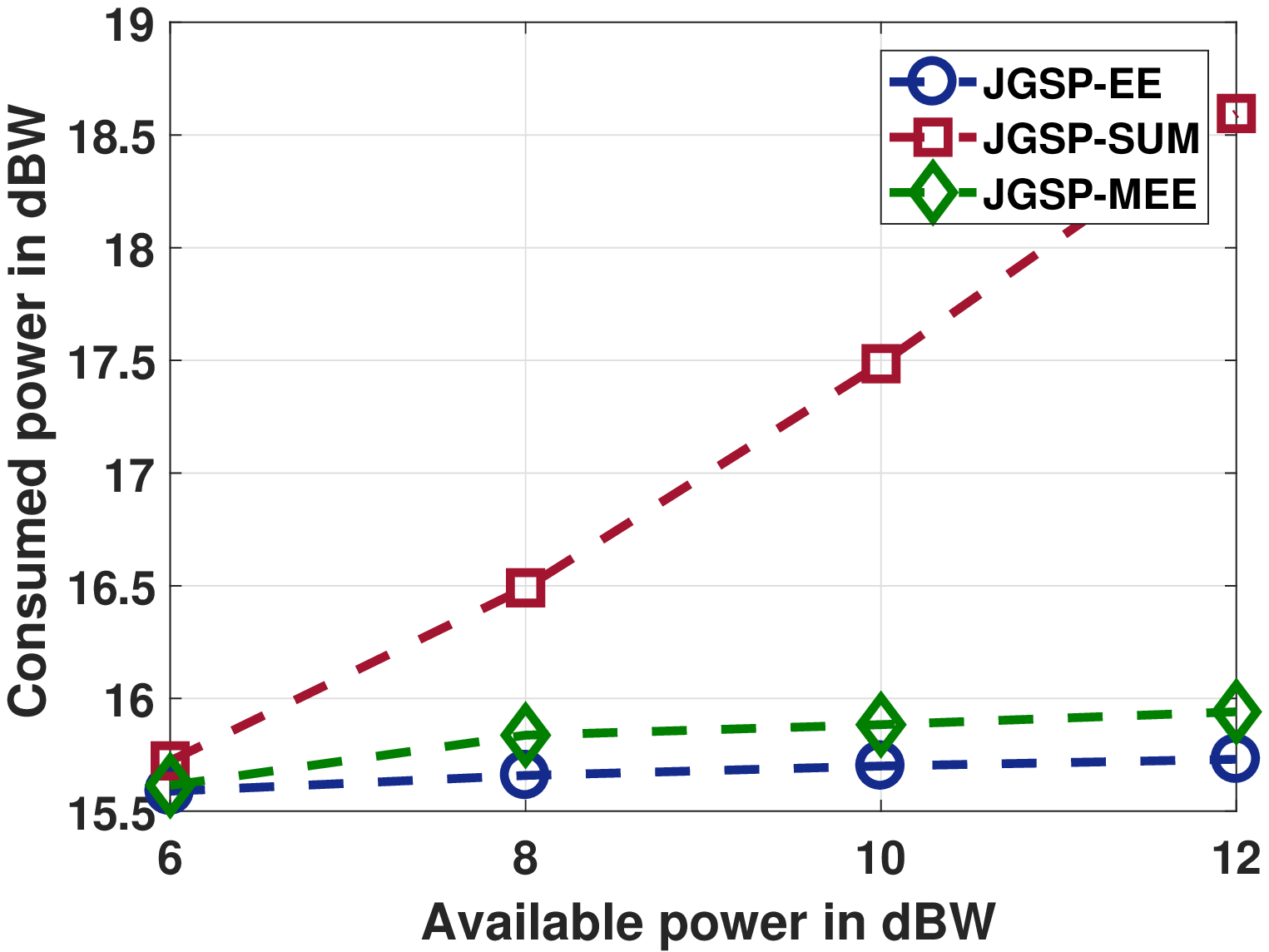}\label{fig:Pow_vs_totpow}}
    \subfloat[]{ \includegraphics[height=6.15cm,width=8cm]{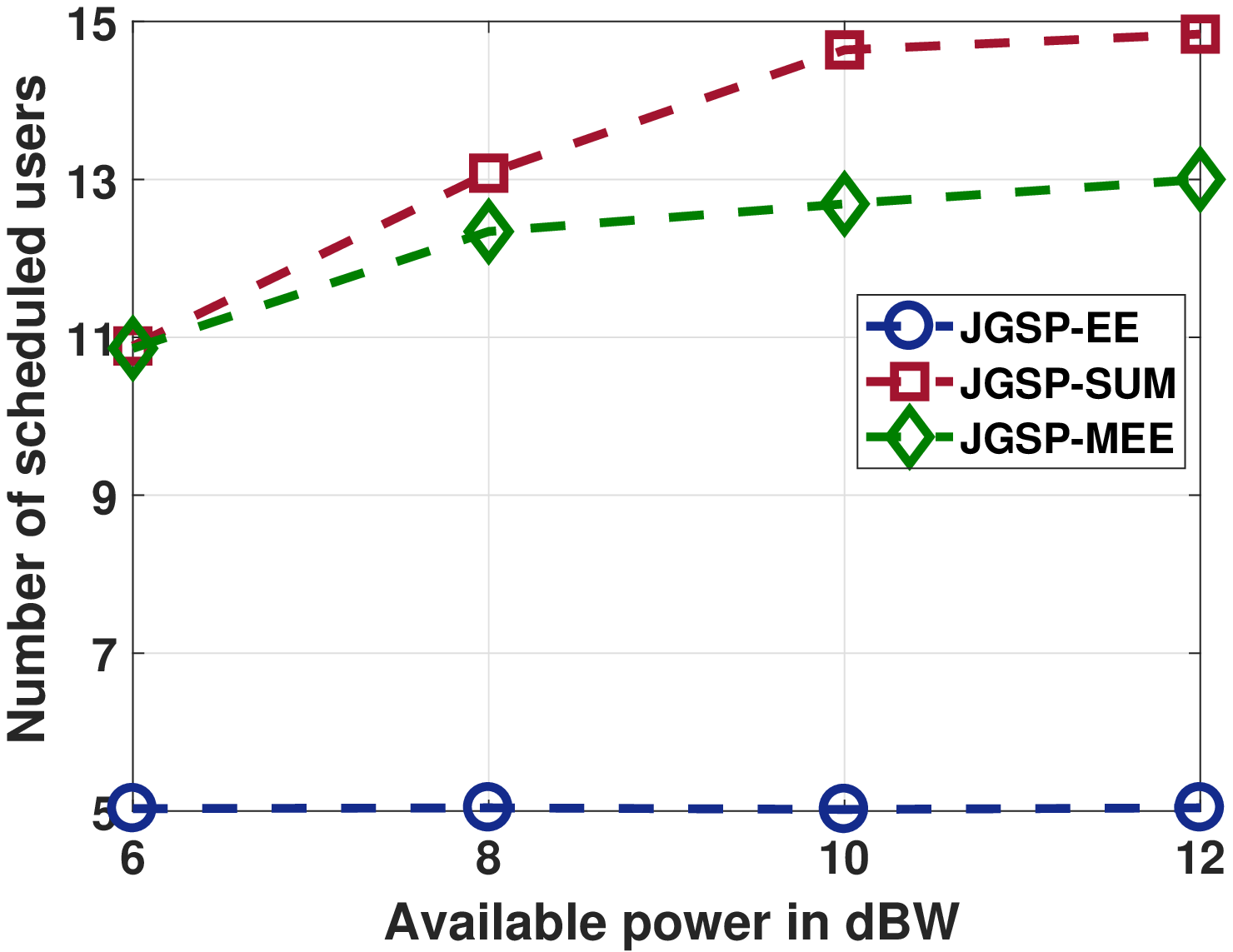}\label{fig:NSU_vs_totpow}} \\
    \subfloat[]{\includegraphics[height=6.15cm,width=8cm]{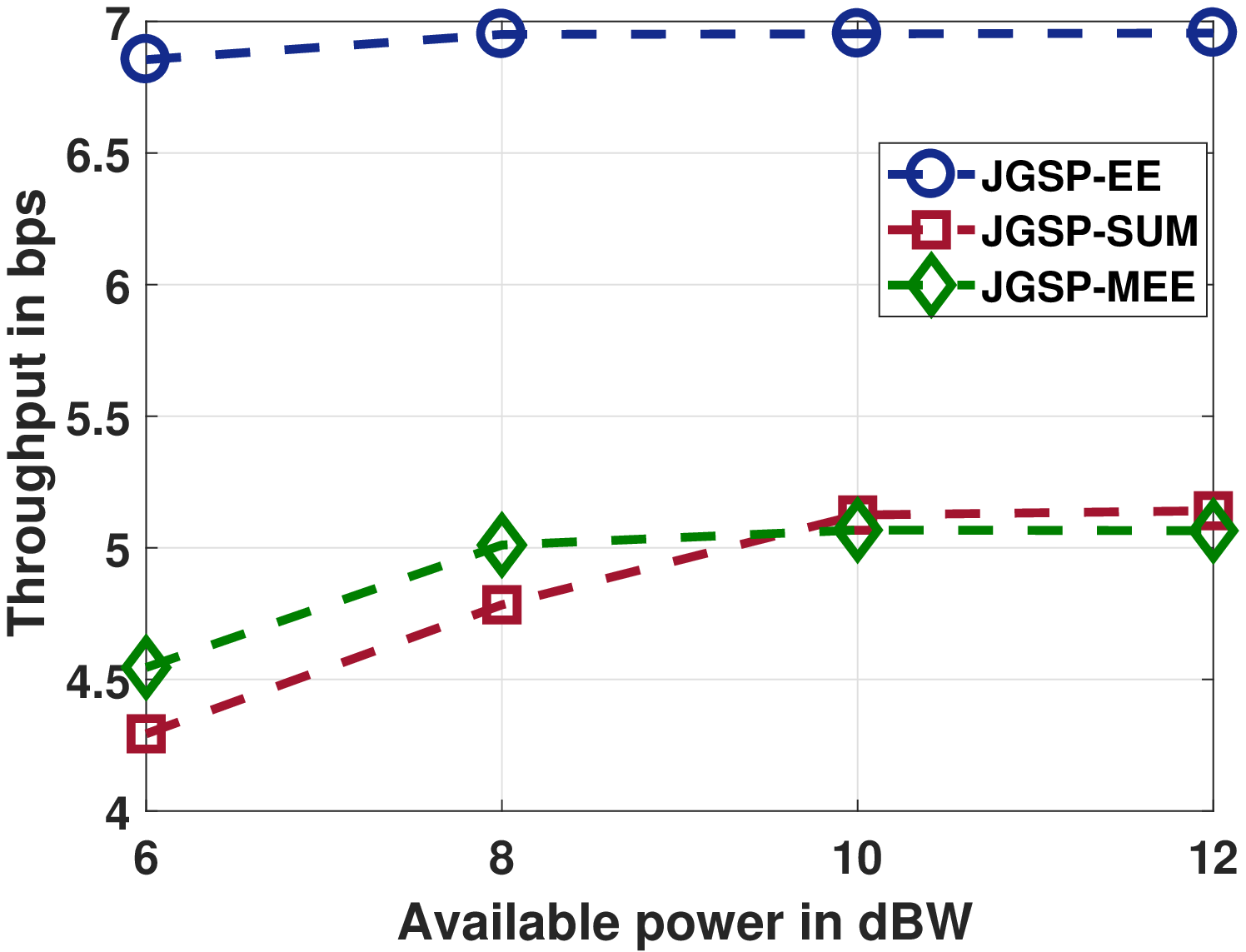}\label{fig:SE_vs_totpow}}
    \subfloat[]{\includegraphics[height=6.15cm,width=8cm]{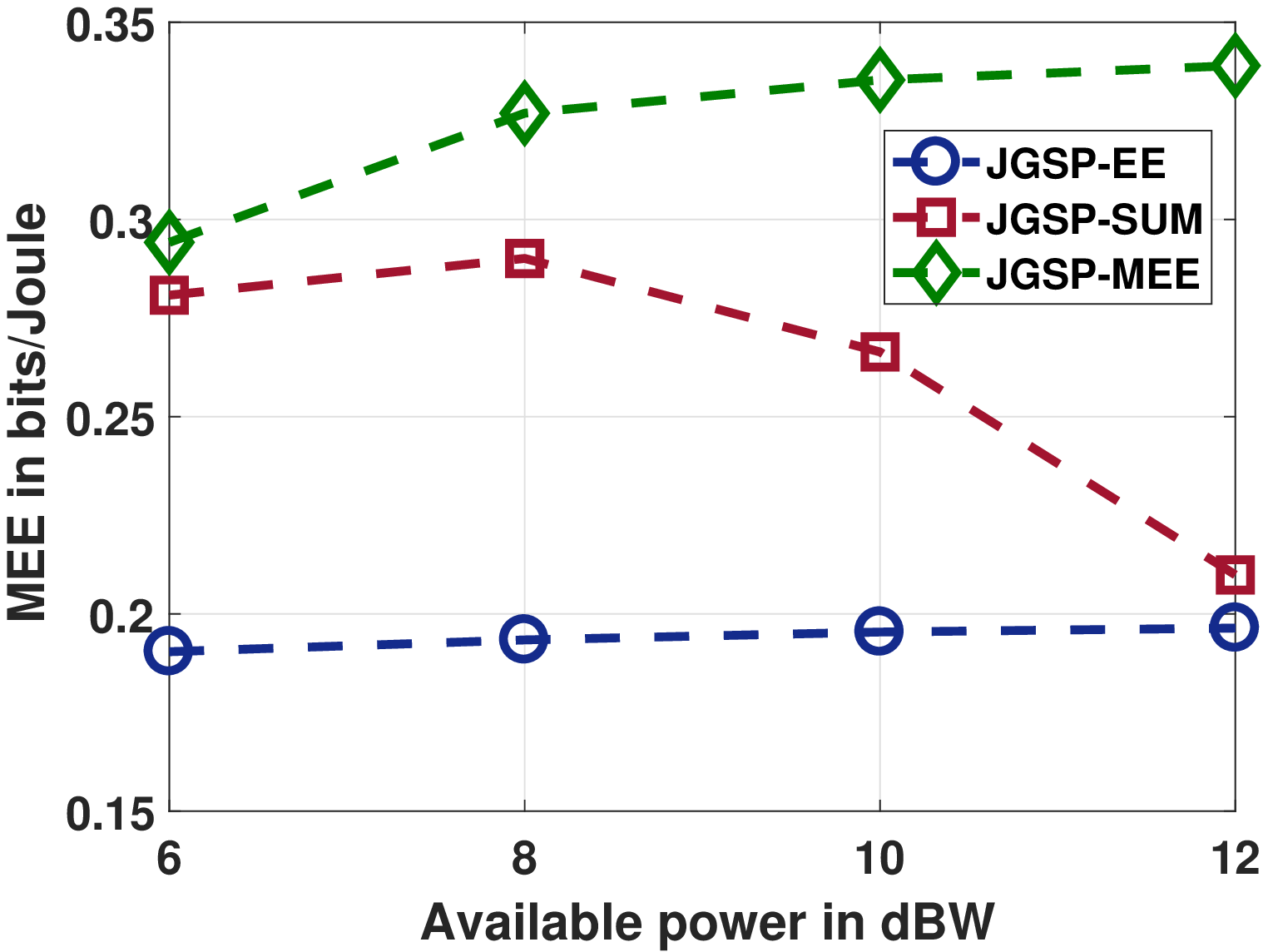}\label{fig:MEE_vs_totpow}}
    \caption{Comparison of proposed algorithms as a function of $\totpow$ varying from 6 to 12 for $\ntxants=5, \ngrps=8$, $\lbrace {\epsilon}_i =1\text{bps/Hz}\rbrace_{j=1}^{\totUsers}$, and $\totUsers=15$ (a) consumed power in dBW versus $\totpow$ (b) number of scheduled users versus $\totpow$ (c) throughput in bps versus $\totpow$  (d) MEE in bits/Joule versus $\totpow$}\label{fig:Perf_vs_totpow}
\end{figure}

In figure \ref{fig:Perf_vs_totpow}, the performance of the proposed algorithms i.e., JGSP-MEE, JGSP-EE, and JGSP-SUM is illustrated as a function of $\totpow$ varying from 6 to 12 in steps of 2 dBW for $M=5$, $\ngrps=8$, $\totUsers=15$ and ${\epsilon}_j=1$ bps/Hz, $\forall j$. 
\subsubsection{Number of scheduled users versus $\totpow$}
In figure~\ref{fig:NSU_vs_totpow}, number of scheduled users is illustrated as function of $\totpow$. By directly maximizing the number of scheduled users, JSP-SUM schedules the maximum number of users compared to JGSP-MEE and JGSP-EE. In the low-available power regime i.e., $\totpow=6$ dBW and 8 dBW, JSP-SUM schedules only few users. However, in the high-available power regime, due to the sufficient power, JSP-SUM schedules almost all the users i.e., 15 users by utilizing all of the power. Unlike JSP-SUM, despite the increased available power, the number of scheduled users in JSP-MEE is saturated to 13 users. This is because scheduling those extra users results in the consumption of huge power which decreases the overall MEE. Moreover, for the available for 8 dBW, JSP-SUM and JSP-MEE schedules almost equal number of users, however, JSP-MEE consumes almost 6.8 dBW less than JSP-SUM. On the contrary, for the reasons mentioned in section~\ref{sec:Perf_N}, JSP-EE schedules only $\ntxants$ despite the availability of power.

\subsubsection{Throughput versus $\totpow$}
In figure~\ref{fig:SE_vs_totpow}, the throughput in bps obtained by JGSP-MEE, JGSP-EE, and JGSP-SUM is illustrated as a function of $\totpow$. Since JSP-MEE and JSP-SUM sacrifice in throughput to schedule more users for the available power, the lower throughput of JSP-MEE and JSP-SUM compared to JSP-EE can be observed in figure~\ref{fig:SE_vs_totpow}. Moreover, the throughput of JSP-EE saturates to 8 bps for available power of 8 dBW as improving throughput further results in the consumption of huge power which results in overall lower EE. On the other hand, in the low-available power regime $\totpow=6$ and 8 dBW, the number of scheduled users (by JSP-SUM) are around 11 and 13 which less than total number of users $\totUsers=15$. In other words, scheduling a higher number of users than 13 requires higher available power than 8 dBW. Therefore, the available-power in this regime is used to improve the minimum throughput of scheduled groups by JSP-MEE which can be observed in figure~\ref{fig:MEE_vs_totpow}. In this high-available power regime i.e., $\totpow \geq 10$ dBW, JSP-SUM uses all of the available power to schedule almost all the users as shown in figure~\ref{fig:MEE_vs_totpow}. On the contrary, despite the availability of the power to schedule all users and/or to improve the throughput, JSP-MEE relatively maintains the same throughput as for the case of $\totpow=8$ dBW since the improvement in throughput leads to consumption of huge power.

\subsubsection{Consumed power versus $\totpow$}
In figure~\ref{fig:Pow_vs_totpow}, the consumed power in Watts by JSP-SUM, JSP-EE and JSP-MEE is illustrated as a function of $\totUsers$. As the JSP-SUM does not optimize power, the inefficient utilization of the available power of JSP-SUM can be observed in figure~\ref{fig:Pow_vs_totpow}. On the contrary, as the EE and MEE objectives include the consumed power in the denominator, the objective values of EE and MEE are decreases inversely for a linear increase in consumed power. Hence, JSP-EE and JSP-MEE utilize power efficiently as shown in figure~\ref{fig:Pow_vs_totpow}. 

\subsubsection{MEE versus $\totpow$}
In figure~\ref{fig:MEE_vs_totpow}, the MEE in bits/Joule obtained by JGSP-MEE, JGSP-EE, and JGSP-SUM is illustrated as a function of $\totpow$. By striking the trade-off among optimizing the number of scheduled users, throughput, and consumed power, JSP-MEE obtains higher MEE compared to JSP-EE and JSP-SUM as shown in figure~\ref{fig:MEE_vs_totpow}. Although JSP-SUM schedules more users than JSP-MEE, it does so by consuming huge power and also by inefficiently utilizing the available power. This results in decreasing in MEE of JSP-SUM with the increase of available power. On the other hand, JSP-MEE schedules also increase the number of users while simultaneously optimizing power and throughput. This results in an overall better MEE of JSP-MEE.
On the contrary, JSP-EE schedules only $\ntxants=5$ users despite the opportunity to schedule more users. Hence, JSP-EE results in the lowest MEE. However, JSP-MEE schedules more users while efficiently utilizing power and throughput.

\section{Conclusions}\label{sec:conclusion}
In this paper, the joint design of user grouping, scheduling, and precoding problem was considered for the message-based multigroup multicast scenario in multiuser MISO downlink channels. In this context, to fully leverage the multicast potential, a novel metric called multicast energy efficiency is considered as a performance metric. Further, this joint design problem is formulated as a structured MINLP problem with the help of Boolean variables addressing the scheduling and grouping, and linear variables addressing the precoding aspect of the design. Noticing the structure in MINLP to be difference-convex/concave, this paper proposed efficient reformulations and relaxations to transform it into structured DC programming problems. Subsequently, the paper proposed CCP based algorithms for MEE and its variants i.e., EE and SUM problems (JSP-MEE, JSP-EE, and JSP-SUM) which are guaranteed to converge to a stationary point for the aforementioned DC problems. Finally, the paper proposed low-complexity procedures to obtain good feasible initial points, critical to the implementation of CCP based algorithms. Through simulations, the paper established the efficacy of the proposed joint techniques and studied the influence of the algorithms on the different parameters namely scheduled users, multicast throughput and consumed power.

\section*{Appendix I}\label{sec:TaylorEXp}
\begin{align}
&\tilde{\mathbb{P}}^{k}\left(\eta_{ij}\right) \triangleq \nonumber \text{ } \eta_{ij}\nabla \mathbb{P}\left(\eta_{ij}^{k-1}\right); \text{ } \tilde{\mathbb{P}}^{k}\left(\delta_j\right) \triangleq \nonumber \text{ }  \delta_{j} \nabla \mathbb{P}\left(\delta_{j}^{k-1}\right) , \\
&f^k\left(\eta_{ij},\Theta_j,t\right)\triangleq \BW \left( 2\dfrac{\left(\eta_{ij}^{k-1}+\Theta_j^{k-1}\right)\left(\eta_{ij}+\Theta_j\right)}{t^{k-1}} -\left(\dfrac{\eta_{ij}^{k-1}+\Theta_j^{k-1}}{{t^{k-1}}}\right)^2t-\dfrac{\eta_{ij}^2}{t}-\dfrac{\Theta_j^2}{t}\right). \nonumber
\end{align}
Similarly, linearization of the concave part of $C_6$, $C_7$ and $C_9$ in $\mathcal{P}_{5}^{\text{MEE}}$ is given by
\begin{align}
&\tilde{\mathcal{G}}_{ij}^{k}(\eta_{ij},\Theta_j) \triangleq -\left[{\eta_{ij}^{k-1}}\right]^2-\left[{\Theta_j^{k-1}}\right]^2 +2\eta_{ij}^{k-1}\eta_{ij}+2\Theta_j^{k-1}\Theta_j, \nonumber \\
&\color{blue}\tilde{\mathcal{K}}_{ij}^{k}(\delta_{j},\Theta_j) \triangleq \left(\delta^{k-1}_{j}+\Theta^{k-1}_j\right)^2-2\begin{bmatrix} \delta^{k-1}_{j}+\Theta^{k-1}_j\\
\delta^{k-1}_{j}+\Theta^{k-1}_j\end{bmatrix}^H\begin{bmatrix} \left(\delta_{j}-\delta^{k-1}_{j}\right)\\
\Theta_{j}-\Theta^{k-1}_{j}\end{bmatrix} \nonumber \\
&\tilde{\mathcal{J}}^k_{ij}(\mathbf{W},\alpha_{ij})  \triangleq -\mathcal{J}_{ij}(\mathbf{W}^{k-1},\alpha_{ij}^{k-1})- \Re \left\{ \nabla^H \mathcal{J}_{ij}(\mathbf{W}^{k-1},\alpha_{ij}^{k-1}) \begin{bmatrix}
    \lbrace {\mathbf{w}_l-\mathbf{w}_l^{k-1}} \rbrace_{l=1}^{\ngrps} \\
    \alpha_{ij}-\alpha^{k-1}_{ij}
    \end{bmatrix} \right\}, \nonumber
\end{align}
where  
\begin{align}\label{eq:Grad}
    &\nabla \mathcal{J}_{ij}({\mathbf{W}^{k-1}},\alpha_{ij}^{k-1}) = \begin{bmatrix}
    \lbrace \dfrac{2{\mathbf{h}_i\mathbf{h}_i^{H}\mathbf{w}_l^{k-1}}}{\alpha_{ij}^{k-1}} \rbrace_{l=1}^{\ngrps} \\
    -\dfrac{\sigma^2+\sum_{l = 1}^{\ngrps}\lvert\mathbf{h}^H_{i}\vecw_{l}^{k-1}\rvert^2}{{\alpha_{ij}^{k-1}}^2}
    \end{bmatrix}.
\end{align}

\section*{Appendix II}\label{sec:appnedix}
\subsection*{DC formulation and CCP based algorithm: SUM}
Applying reformulations and relaxations proposed in Section~\ref{sec:MEE}, the problem $\mathcal{P}_{1}^{\text{SUM}}$ is reformulated into DC problem as,
 \begin{align}\label{eq:SUM_Bin_form2}
    \mathcal{P}_{2}^{\text{SUM}}:&\text{} \max_{\mathbf{W},\boldsymbol{\Theta,\eta,\delta} } \text{ }  \sum_{i=1}^{\totUsers}\sum_{j=1}^{\ngrps} \eta_{ij}- \Omega_3 \norm{\sum_{j=1}^{\ngrps}\delta_{j}-\ntxants}^2 +\lambda_4 \sum_{j=1}^{\ngrps}\sum_{i=1}^{\totUsers}{\mathbb{P}}\left(\eta_{ij}\right) +\sum_{j=1}^{\ngrps} \lambda_5 {\mathbb{P}}\left(\delta_j\right)  \\ 
    \text{ s.t. } 
    &C_1:\text{ } 0 \leq \eta_{ij} \leq 1, \text{ } \forall i, \text{ }\forall j,
    \hspace{5cm}C_{4}:\text{ } 0 \leq \delta_{j} \leq 1, \forall j,\nonumber \\
    &C_{7}:\text{ }\sum_{l\neq i}|\mathbf{h}_{i}^H\mathbf{w}_l|^2+\sigma^2 \leq \dfrac{\sum_{l=1}^{\ngrps}|\mathbf{h}_{i}^H\mathbf{w}_l|^2+\sigma^2}{1+\eta_{ij}\epsilon_j}, \text{ }\forall i, \text{ }\forall j, \hspace{0.4cm}C_2, C_{3}, C_5 \text{ and } C_6 \text{ in } \eqref{eq:SUM_Bin_form1}, \nonumber
    \end{align} 
where ${\lambda}_5, {\lambda}_6$ and ${\Omega}_3$ are the penalty parameters. 

The convexified problem to be solved as part of JGSP-SUM (CCP based algorithm applied to the DC problem $\mathcal{P}_{2}^{\text{SUM}}$) algorithm at iteration $k$ is:
\begin{align}\label{eq:SUM_CCP_joint_convx}
  \mathcal{P}_{3}^{\text{SUM}}:\text{}\max_{\mathbf{W},\boldsymbol{\Theta,\eta,\delta}} \text{ } &\sum_{i=1}^{\totUsers}\sum_{j=1}^{\ngrps} \eta_{ij}- \Omega_3 \norm{\sum_{j=1}^{\ngrps}\delta_{j}-\ntxants}^2   +{\lambda}_4 \sum_{j=1}^{\ngrps} \sum_{i=1}^{\totUsers} \tilde{\mathbb{P}}^k\left(\eta_{ij}\right)+\sum_{j=1}^{\ngrps} {\lambda}_5 \tilde{\mathbb{P}}^k\left(\delta_j\right)  \\ 
    \text{ s.t. } 
    C_1, C_2, &C_3, C_5 \text{ to } C_6 \text{ in } \eqref{eq:SUM_Bin_form2},  \hspace{0.3cm}
    C_{7}:\sum_{l\neq i}|\mathbf{h}_{i}^H\mathbf{w}_l|^2+\sigma^2 \leq \mathcal{I}^{k}\left(\mathbf{W},\eta_{ij}\right), \text{ }\forall i, \text{ }\forall j, \nonumber 
\end{align} 
where
$\mathcal{I}^k\left(\mathbf{W},\eta_{ij}\right)=\dfrac{\sum_{l=1}^{\ngrps}|\mathbf{h}_{i}^H\mathbf{w}^{k-1}_l|^2+\sigma^2}{1+\eta^{k-1}_{ij}\epsilon_j} + \nonumber \\
\Re \left\{ \begin{bmatrix}
    \lbrace \dfrac{2{\mathbf{h}_i\mathbf{h}_i^{H}\mathbf{w}_l^{k-1}}}{1+\eta_{ij}^{k-1}\epsilon_j} \rbrace_{l=1}^{\ngrps} 
    -\epsilon_j\dfrac{\sigma^2+\sum_{l = 1}^{\ngrps}\lvert\mathbf{h}^H_{i}\vecw_{l}^{k-1}\rvert^2}{{\left(1+\eta_{ij}^{k-1}\epsilon_j\right)}^2}
    \end{bmatrix}^{H}\begin{bmatrix}
    \lbrace {\mathbf{w}_l-\mathbf{w}_l^{k-1}} \rbrace_{l=1}^{\ngrps} \\
    {\eta_{ij}-\eta^{k-1}_{ij}}
    \end{bmatrix} \right\}.$
    
Letting $\mathcal{P}_{3}^{\text{SUM}}\left(k\right)+$ be the objective value of the problem $\mathcal{P}_{3}^{\text{SUM}}$ at  iteration $k$, the pseudo code of JGSP-EE-SR for the joint design problem is given in algorithm~\ref{alg:JGSP_EE_SUM}.
\begin{algorithm}
 \caption{JGSP-SUM}
 \label{alg:JGSP_EE_SUM}
 \begin{algorithmic}[]
 \State{\textbf{Input}: $\mathbf{H},\left[\epsilon_1,\hdots,\epsilon_{\totUsers}\right],\totpow,\Delta, \mathbf{W}^0, \boldsymbol{\delta}^0,\boldsymbol{\Theta}^0,\boldsymbol{\eta}^0$, $\lambda_4=0, \lambda_5=0, \Omega_3=0, k=1$}; \\
 \textbf{Output}: $\mathbf{W},\boldsymbol{\eta}$
 \While{$|\mathcal{P}_{3}^{\text{SUM}}\left(k\right)-\mathcal{P}_{3}^{\text{SUM}}\left(k-1\right)|\geq \Delta$}
    \State \textbf{Convexification:} Convexify the problem \eqref{eq:SUM_CCP_joint_convx}
    \State \textbf{Optimization}: Update $\left({\mathbf{W}}, \boldsymbol{\eta,\delta,\Theta}\right)^{k}$ by solving $\mathcal{P}_{3}^{\text{SUM}}$
    \State \textbf{Update :} $\mathcal{P}_{3}^{\text{SUM}}\left(k\right), {\lambda}_4,{\lambda}_5,{\Omega}_3, k$
 \EndWhile
\end{algorithmic}
\end{algorithm}  

\bibliographystyle{IEEEtran}
\bibliography{IEEEabrv,bibJournalList,EE_MGMC}



\end{document}